\documentclass[
  journal=largetwo,
  manuscript=article-type,
  year=2020,
  volume=37,
]{cup-journal}

\usepackage{amsmath}
\usepackage[nopatch]{microtype}
\usepackage{booktabs}
\usepackage{multirow}
\usepackage{subcaption}
\usepackage{mwe}

\usepackage{journal_abbr_mac}

\newcommand{\comment}[1]{}
\def\Msun{\>{\rm M_{\odot}}}

\DeclareUnicodeCharacter{2212}{-}

\definecolor{ref@color}{RGB}{0,76,153}

\title{Stealing galaxies from galaxy clusters}

\author{M. Smole}
\affiliation{Astronomical Observatory, Volgina 7, 11060 Belgrade, Serbia}
\email[M. Smole]{msmole@aob.rs}

\author{M. Micic}
\affiliation{Astronomical Observatory, Volgina 7, 11060 Belgrade, Serbia}

\author{A. Mitra\v{s}inovi{\'c}}
\affiliation{Astronomical Observatory, Volgina 7, 11060 Belgrade, Serbia}


\begin{document}

\begin{abstract}

We investigate galaxy groups that reside in the field but have been previously processed by galaxy clusters. Observationally, they would appear to have the same properties as regular field groups at first glance. However, one would expect to find quantifiable differences in processed groups as dynamical interactions within clusters perturb them. 
We use IllustrisTNG300 simulation to statistically quantify that processed groups of galaxies show different properties compared to regular field groups.

Our analysis encompasses a broad range of groups with total masses between
$8 \times 10^{11} \Msun$ and $7 \times 10^{13} \Msun$.
We distinguish between processed groups that passed through a galaxy cluster and capture more
galaxies, referred to as thief groups, and groups that did not capture any new members,
referred to as non-thief groups.
The employed statistical tools show that thief groups are generally less compact and contain more members, while non-thief groups seem to have the same properties as the field groups which makes them indistinguishable.

\end{abstract}

\section{Introduction}

According to the $\Lambda$CDM standard cosmological model, our Universe is composed of dark energy ($\sim 68.3 \%$) and cold dark matter ($\sim 26.5 \%$), while the remaining goes to baryonic matter and radiation (see \cite{Planck_Collaboration} for precise values of cosmological parameters). $\Lambda$CDM model relies on the Cosmological principle, which indicates that the Universe is isotropic and homogeneous on large scales ($>100\;$Mpc). On smaller scales, the Universe shows an inhomogeneous structure, which is the result of primordial quantum fluctuations. During the inflation era, quantum fluctuations grow to macroscopic scales, and gravitational collapse leads to the formation of the first structures in the Universe \citep[e.g.,][]{Peebles1980, Peebles1993, Padmanabhan1993}. Dark matter (DM) halos were the first objects to form, followed by the gravitational collapse of baryonic matter in their centres and the formation of the first galaxies. Higher-density regions become more dense as gravitational pull attracts galaxies together, and they first form groups that later grow into galaxy clusters containing thousands of members.

At present-day redshift, $z=0$, the large-scale structure of the Universe shows the Cosmic Web, the interconnected network composed of low-density regions called voids and DM filaments at whose intersections reside galaxy clusters, the highest-density regions in the Universe. Galaxy clusters grow by constant accretion of gas and DM, as well as galaxies, groups of galaxies, and smaller clusters that flow along DM filaments.

The $\Lambda$CDM standard cosmological model posits that DM halos and their associated galaxies grow hierarchically, primarily through mergers with other galaxies. Moreover, individual galaxy evolution depends on the type of environment in which the galaxy resides, which should be intuitively understood as a consequence of the hierarchical growth paradigm. In other words, galaxies in low-density regions (such as voids) experience mergers very rarely, so their growth occurs mainly due to the inflow of gas through the DM filaments. Such galaxies are generally gas-rich spiral galaxies with lower masses and higher star formation. In contrast, in high-density environments, galaxy mergers are much more frequent, which influences the galaxy's evolution, leading to the formation of red elliptical galaxies with no star formation. This dichotomy of galactic morphology that depends on the environment is commonly known as the morphology-density relation \citep{dressler1980tsigmadata, dressler1980tsigma}.

Further, consecutive mergers of elliptical galaxies form massive elliptical galaxies (cDs) at the centres of galaxy clusters, the most massive galaxies ever discovered (see \cite{Micic2013} for a review). Alongside a higher likelihood of galaxy mergers in clusters, tidal interactions during galaxy flybys, a type of non-merger close interaction \citep{sinha2012}, can also strongly influence galaxy morphology \citep[e.g.,][]{Pettitt+Wadsley2018, Mitrasinovic+Micic2023}. Galaxy flybys play an important role in the evolution of galaxies in rich galaxy clusters where, due to high relative velocities between galaxies, they can be much more frequent than mergers and often outnumber them by an order of magnitude \citep{shan2019}.

Interactions between gas in the interstellar medium in galaxies and the intercluster medium can lead to ram pressure stripping (RPS). This process efficiently removes the galactic gas, finally quenching its star formation. RPS can be responsible for turning the evolution of spiral galaxies into lenticular (S0) galaxies without star formation. Recent studies (\cite{Lopes}; \cite{Piraino-Cerda} and references therein) have shown evidence of RPS in galaxies that reside on the outskirts of the cluster, up to $\sim 5\; R_{\textrm{vir}}$.
Similarly, studies based on the HI catalogue of galaxies observed by 
Widefield ASKAP L-band Legacy All-sky Blind surveY (WALLABY, \cite{Koribalski2012};
\cite{Koribalski2020}) have shown evidence of gas removal and RPS in group and cluster
environments (\cite{Reynolds2022}; \cite{Lin2023}; \cite{Holwerda2023,Holwerda2025}).

Galaxy And Mass Assembly (GAMA) survey provides an extensive research 
of the properties of galaxy groups (\cite{Mata}; \cite{Sotillo-Ramos}; \cite{Banks}; \cite{Riggs};  \cite{Dev}; \cite{Lombilla}).
The significant fraction of starburst and passive galaxies in groups around galaxy clusters suggests that RPS can quench star formation even before groups become cluster members, often referred to as pre-processing of galaxy groups. Before quenching, the stripped gas can form stars outside the galactic centre, leading to the formation of jellyfish galaxies (see, e.g., \cite{Ebeling2014}; \cite{Fumagalli}; \cite{Poggianti2017}). The highest fraction of jellyfish galaxies is found in galaxy clusters, indicating that RPS plays an important role in the evolution of cluster galaxies (\cite{Vulcani2022}; \cite{Piraino-Cerda}).

Compact elliptical galaxies (cEs) are another example of a rare galaxy class, with a mass typically above $10^{9}\Msun$ and a radius less than 1 kpc. Most cEs are found in the cores of galaxy clusters, usually
orbiting around massive host galaxies \citep{Huxor2011}. This suggests that cEs are formed by tidal interactions, where a more massive galaxy strips matter from the outskirts of its satellite galaxy, leaving the high-density stellar core. However, cEs have also been found in isolation (\cite{Huxor2013}; \cite{Paudel2014}). The different observed properties of isolated and host-associated cEs \citep{Kim2020}
suggest separate formation pathways for cEs in different environments. Isolated cEs follow the mass-metallicity relation of ellipticals, whereas cEs orbiting a more massive host have higher metallicities and appear redder, smaller, and older. This was supported by a recent study \citep{Deeley} that found that $\sim30$ percent of cEs are formed via the stripping of a spiral galaxy by a host galaxy, while the remaining $\sim70$ percent are formed by continuous growth in isolation. \citet{Chilingarian} proposed an alternative scenario in which isolated cEs might be processed by a galaxy cluster and then ejected by three-body encounters. Galaxies receiving kick velocities that are high enough can leave the cluster environment without being accreted by massive hosts. \citet{Deeley} did not find an example of a gravitationally ejected cEs in the IllustrisTNG50 simulation box (\cite{Springel18}; \cite{Naiman}; \cite{Pillepich}; \cite{Nelson}; \cite{Marinacci}). However, a larger simulation box might be needed to identify such rare objects, considering that the TNG50 simulation box features only one massive Virgo-like cluster, making the sample size of cluster galaxies rather small.

Different studies investigated galaxies that had resided inside a cluster, moved to distances up to several virial radii of the host cluster, and then eventually fell back into the cluster, called 'backsplash' galaxies (e.g., \cite{Borrow}; \cite{Ruiz}). Studying this class of galaxies can provide important insights into galaxy evolution in various environments. \citet{Mitrasinovic2023} have shown an example from IllustrisTNG300 of an isolated dark matter-poor galaxy that had been processed by a galaxy cluster and then left the cluster environment, spending the last $\sim$ 2 Gyr in isolation. At $z=0$, the galaxy has a stellar mass of $M_\star=6.8 \times 10^9 \Msun$ and a stellar half-mass radius of $R_{0.5,\star}=2.45$ kpc. Even though the galaxy does not satisfy the observational compactness criterion, such as one given by \citet{Barro+2013ApJ...765..104B}, it represents an example of a fairly compact galaxy that has managed to escape the cluster. 

Clearly, individual galaxies are known to escape the gravitational potential of galaxy clusters through dynamical interactions or high-velocity encounters. By extension, it is plausible, albeit less common, for entire groups of galaxies to undergo a similar process. Backsplash galaxies, or those that were completely ejected from galaxy clusters, are kinematically distinct from galaxies in the field or their first cluster infall \citep{Gill+2005} due to their prior interaction with the cluster as a whole and individual member galaxies. Similarly, it is reasonable to assume that escaped groups may also exhibit unique characteristics imprinted by their prior interaction with the cluster, unlike field groups that have never been part of a cluster environment. By studying these escaped groups, we can probe the long-term effects of the cluster environment on galaxy and group evolution and assess how much they differ from unprocessed groups. The focus of our work is to statistically quantify that processed groups of galaxies show different properties compared to regular field groups.

Here, we investigate how galaxy groups can be processed by a cluster using IllustrisTNG300 cosmological simulation. We study groups of galaxies that form in isolation and then fall together inside a cluster. Typically, after one pericentric passage, these groups escape the cluster, sometimes accreting more members in the cluster phase. We explore the possible effects of a cluster passage on those galaxy groups as well as on individual group members. In Section~\ref{method}, we describe the data set and methodology used in this work. We present and discuss our findings in Section~\ref{results} and draw conclusions in Section~\ref{conclusion}.

\section{Methods}
\label{method}

We use IllustrisTNG cosmological hydrodynamical simulations of galaxy formation
to investigate galaxy groups in different environments (\cite{Springel18}; \cite{Naiman}; \cite{Pillepich}; \cite{Nelson}; \cite{Marinacci}). Simulations within the IllustrisTNG project were performed using the \textsc{arepo} code (\cite{Springel10}) and the \citet{planck} cosmological parameters.
We use the simulation with the largest cosmological box, TNG300
$(\sim 300~ \rm{Mpc})^{3}$, which offers the largest sample of galaxy clusters.
TNG300 simulation has a mass resolution of $5.9 \times 10^7 \Msun$ for DM particles and $1.1 \times 10^7 \Msun$ for baryonic components. 

We place a special interest in the unusual evolutionary path of groups that escaped galaxy clusters and potentially accreted more galaxies from the cluster.

\subsection{Sample selection}

Our analysis includes a wide variety of groups with total masses ranging from
$8 \times 10^{11} \Msun$ to $7 \times 10^{13} \Msun$. 
The lower mass limit is imposed by the resolution of the simulation, ensuring that individual galaxies are resolved with a sufficient number of particles.
The upper limit is selected to exclude galaxy cluster-sized halos, which are typically characterised by masses $>8\times10^{13}\Msun$ (\cite{Paul}).
At redshift $z=0$, the TNG300 simulation box contains 48072 group halos in the given mass range, hosting at least two galaxies.
Here, group halo refers to the gravitationally bound structure identified with the friends-of-friends (FoF) group finding algorithm. In contrast, galaxies refer to the Subhalo fields, identified with the Subfind algorithm, which includes the baryonic component.
Although isolated FoF halos typically have only one galaxy, group and cluster FoF halos can
host ten to thousands of galaxies. Thus, the group mass provides information about the environment in which galaxies reside. 

In order to analyse the evolutionary paths of our sample groups, we trace back the evolutionary tree of each galaxy within the group. To ensure that each individual galaxy is well-resolved, we use the selection criterion $M_\star> 10^8 \Msun$, and we only include structures of cosmological origin that have SubhaloFlag = 1. This filter excludes fragments and clumps that were formed
via internal galaxy processes, such as disk instabilities, and falsely identified as galaxies by the Subfind algorithm (SubhaloFlag = 0).

Even though at $z=0$ our sample groups reside in group-scale halos, by
tracing back the group mass history of individual galaxies, we find that a percentage of galaxies have spent some time in galaxy clusters. 
Galaxies are considered to be part of a cluster if they reside in
FoF halos with mass $>8\times10^{13}\Msun$ (\cite{Paul}). We note that this is a strict definition of a galaxy cluster, since proto-clusters at high redshifts can have lower masses.
Galaxies have left the cluster environment when the mass of their FoF halos drops below this limit.
Typically, snapshots when a galaxy enters or leaves a cluster environment are characterised
by steep increases or drops in host FoF halo mass, usually by two orders of magnitude.
At the first snapshot following the cluster escape, galaxies from our sample reside at a mean distance of $r/R_{\textrm{vir}}\geq3$.

Those galaxies represent rare examples of objects that have managed to escape a galaxy cluster gravitational pull and, in some cases, as we will see below, captured new galaxies from a galaxy cluster on their way out, thus stealing the galaxies from galaxy clusters.

We divided our sample into groups that evolved in isolation,
hereafter referred to as 'field groups', and groups that were a part
of a galaxy cluster at any time during their history, referred to as 'cluster groups'. 
After applying the selection criteria and excluding group halos with one single member,
our final sample contains 32983 field groups and 49 cluster groups. Thus, the cluster groups are $<0.2\%$ of the entire sample.

Cluster groups can be further divided into two sub-samples: groups that passed through a galaxy cluster and captured more galaxies from that cluster, referred to as 'thief groups' (23 cluster groups), and groups that did not capture any new members, referred to as 'non-thief groups' (26 cluster groups).

\subsection{Group compactness parameter}

To estimate the compactness of the group, we calculate the mean distance between its members, weighted by the total group mass. 
More precisely, we measure the mass-weighted mean separation between individual galaxies and the group's centre of mass, calculated as the position of the particle with the minimum gravitational potential energy
('GroupPos' field in the Group catalogue). We define the group compactness parameter as follows:

\begin{equation}
\label{dmean}
    \langle d_{\mathrm{w}} \rangle = \langle \frac{d_{i}\times M_{i}}{M_{\mathrm{tot}}} \rangle.
\end{equation}

\noindent where $d_{i}$ indicates the separation of the individual galaxy from the group's centre of mass, 
$M_{i}$ is the total mass of the individual galaxy, and $M_{\textrm{tot}}$ is the total group mass. Defined this way, higher values of the group compactness parameter imply, perhaps counterintuitively, that the group is less compact. 

In Section \ref{group_compac}, we compare the compactness of field and cluster groups in order to explore how the cluster passage might have influenced the group. 
For a more detailed analysis, we employ the statistical tests described in the following section. 

We also explore details of group cluster passage, with a special focus on the unusual evolutionary path of thief groups.

\subsection{Statistical tests}
\label{stat_test_metod}

The Kolmogorov-Smirnov (KS) test \citep{ks_test} is a commonly used tool to test if
two one-dimensional samples came from the same distribution. 
The KS test is a non-parametric test which calculates the maximum difference
between the two distributions, and the associated p-value.
The p-value represents the probability that the two distributions would be as different as observed
if they were randomly drawn from the same unknown distribution, which serves as the null hypothesis.
Low p-values indicate that the null hypothesis can be rejected, thus indicating that two samples come from different distributions. The significance level of 0.05 is commonly used to indicate statistical significance. Higher p-values indicate that two samples come from the same distribution.

Different approaches can be used to extend the KS test to two-dimensional data samples. In this work, we explore the Cram\'er test, the Peacock test, and the Kernel Consistent Density Equality Test with Mixed Data Types, which we describe in more detail below.

Peacock test \citep{peacock} is a two-dimensional generalization of the KS test, available as \texttt{Peacock.test} package in R.
Using Monte Carlo simulations to test significance levels for two-dimensional distributions,
\cite{peacock} provided the empirical formula to calculate significance levels.
If $D$ is the maximum absolute difference between two distributions, $Z_{\textrm{n}}$ statistics can be defined as $Z_{\textrm{n}}=nD$, with $n=\sqrt{n_{1}n_{2}/(n_{1}+n_{2})}$, where $n_{1,2}$ are the sizes of two samples. The asymptotic value of the Z statistics can be fitted as:

\begin{equation}
1-Z_{\textbf{n}}/Z_{\infty}=0.53~n^{-0.9}.
\end{equation}

\noindent For sufficiently large samples ($n\ge10$), \citet{peacock} gives an analytical approximation for significance levels:

\begin{equation}
    p(>Z_{\infty})=2~\mathrm{exp}~ [-2~(Z_{\infty}-0.5)^{2}],
    \label{peacock_p}
\end{equation}
 and argues that this probability might be too large by, at very most, a factor of $\sim1.5$ and generally within $\sim1.1$.

Another test we use is the non-parametric Cram\'er test for the two-sample problems \citep{cramer} available as an R package \texttt{cramer}. 
Monte Carlo bootstrap methods and eigenvalue methods are available for the calculation of the critical value and for estimating the p-value.
In this work, we use the bootstrap method with 1000 bootstrap-replicates, the default value of this parameter.

\citet{nptest} proposed another non-parametric test for equality of distributions,
Kernel Consistent Density Equality Test with Mixed Data Types, implemented in R as \texttt{npdeneqtest} function included in \texttt{np} package.
We will refer to this test as the npdeneqtest hereafter.
In this approach, the distributions are smoothed by the smoothing parameters chosen via least-squares cross-validation.
The test computes the integrated squared density difference between the estimated densities of two samples.
Again, the test p-value is delivered using bootstrap methods. For this method, we do 1000 bootstrap-replicates, the same as for the Cram\'er test.

\section{Results}
\label{results}

\subsection{Group compactness}
\label{group_compac}

  \begin{figure*}[hbt!]
   \resizebox{\hsize}{!}
            {\includegraphics[width=.45\textwidth]{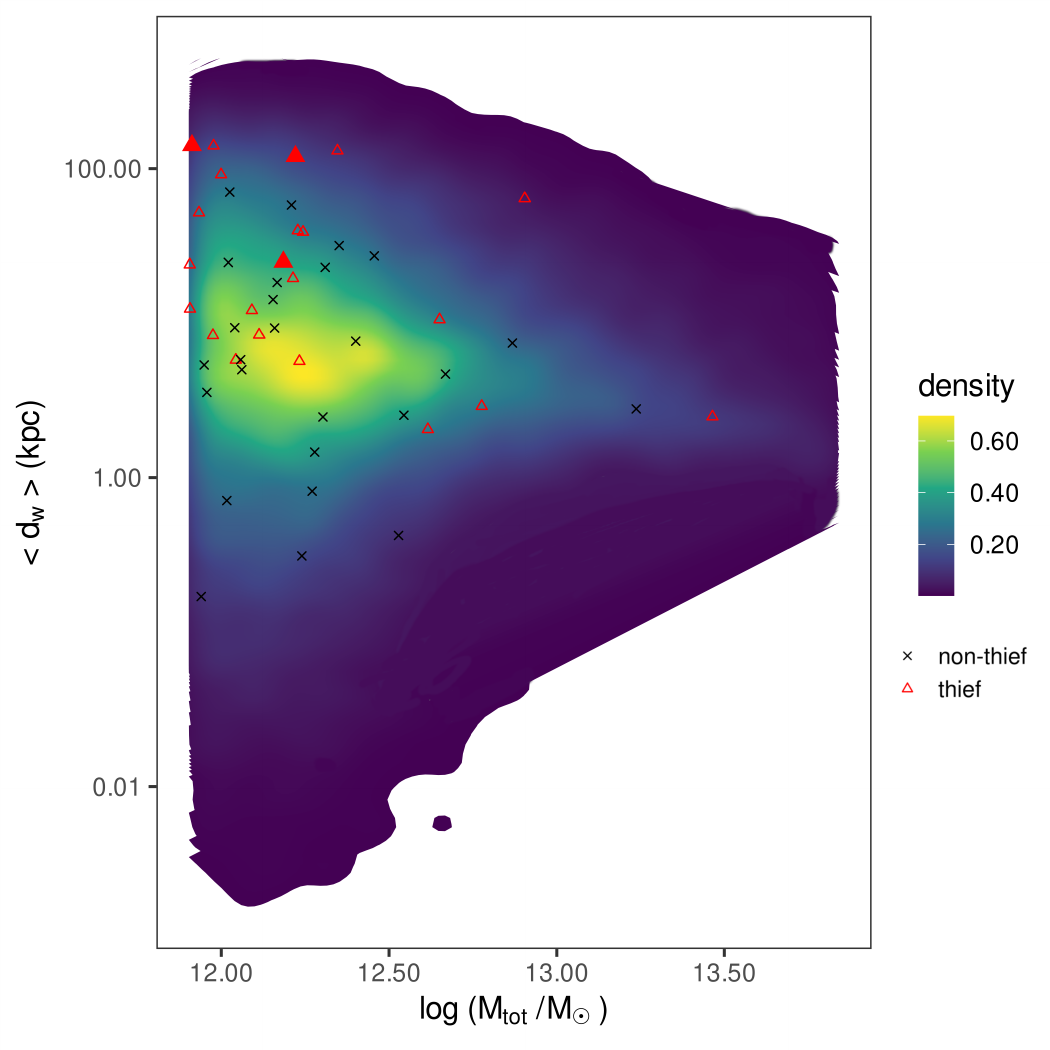}\hfill
\includegraphics[width=.45\textwidth]{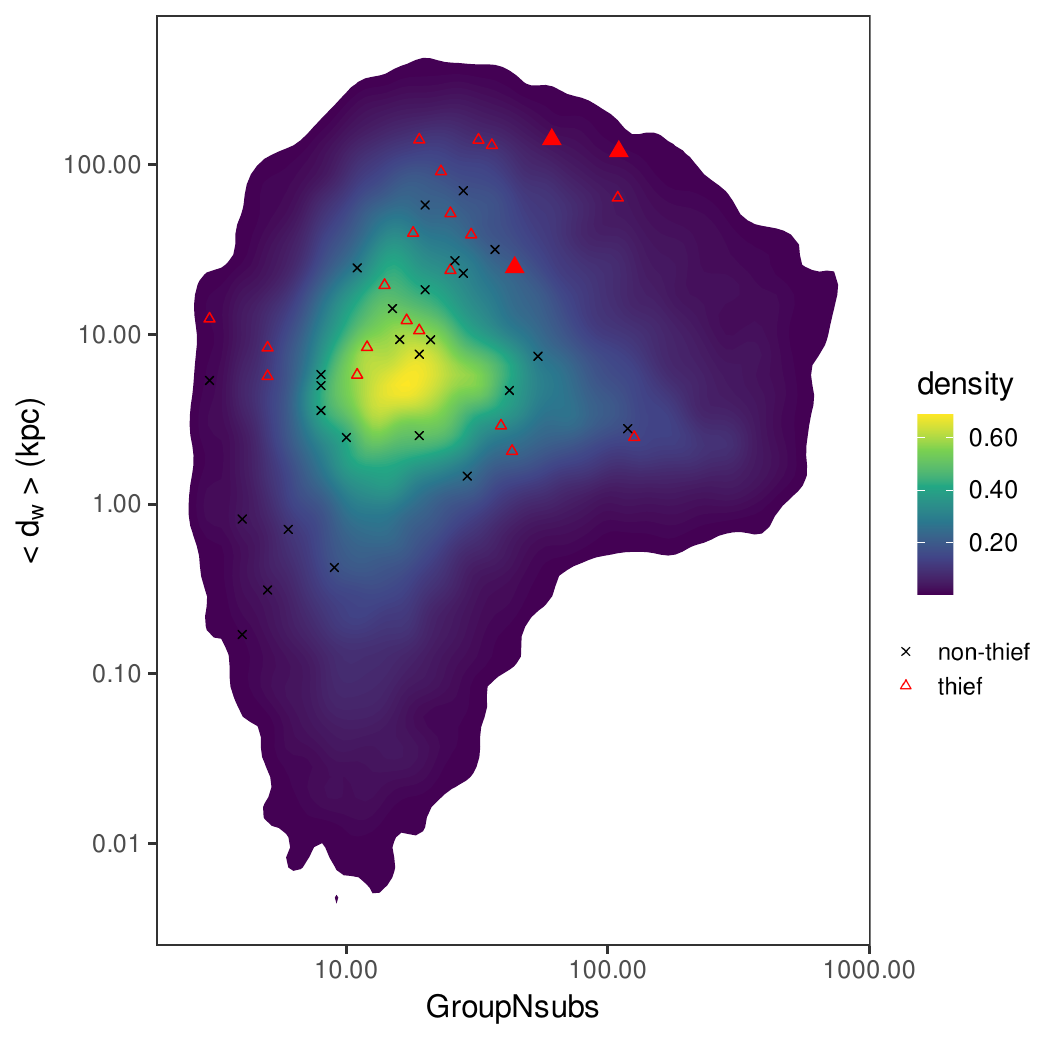}}
    \caption{Galaxy compactness parameter $\langle d_{\mathrm{w}} \rangle$
as a function of the total group mass (left panel) and the total number of group members (right panel).
The density plot shows the distribution of field groups. 
Cluster groups are represented with black crosses (non-thief groups) and 
red triangles (thief groups).
}
    \label{kompaktnost}
\end{figure*}

\begin{table*}[]
\begin{threeparttable}
\caption{p-values from statistical tests}
\label{stat_test}
\begin{tabular}{cccccc}
\multicolumn{2}{c}{}                                                & field vs cluster    & field vs thief       & field vs non-thief & thief vs non-thief  \\ \hline
$M_{\mathrm{tot}}$ - $\langle d_{\mathrm{w}} \rangle$ & cramer      & $1.99\times10^{-3}$ & 0                    & 0.91               & $4.00\times10^{-3}$ \\
                                                      & npdeneqtest & 0.02 & $2.22\times10^{-16}$ & 0.69               & 0.07                \\
\multicolumn{1}{l}{}                                  & peacock     & $1.91\times10^{-3}$ & $2.25\times10^{-4}$ & 0.57               & 0.14                \\ \hline
$N_{\mathrm{gal}}$ - $\langle d_{\mathrm{w}} \rangle$ & cramer      & 0.03 & $1.99\times10^{-3}$                    & 0.18               & $4.45\times10^{-3}$ \\
\multicolumn{1}{l}{}                                  & npdeneqtest & 0.01 & $2.22\times10^{-16}$ & 0.43               & 0.06                \\
                                                      & peacock     & $6.77\times10^{-3}$ & $3.35\times10^{-3}$  & 0.21              & 0.15              
\end{tabular}
\end{threeparttable}
\end{table*}

Figure~\ref{kompaktnost} shows the group compactness parameter $\langle d_{\mathrm{w}} \rangle$
as a function of the total number of group members\footnote{The total number of Subfind groups within the FoF group, including subhalos below our imposed mass limit.} (right panel) and the total group mass (left panel).
The distribution of field groups is shown as a density plot, where the colour bar indicates the estimated 2D density of the data points.
Cluster groups are represented as individual triangles and cross points.  
Black crosses show cluster groups that entered a cluster as already formed groups and, after one or more orbits, leave the cluster without increasing the number of members, thus non-thief groups.
Red triangles represent thief groups, a sub-sample of cluster groups that have gravitationally captured
additional galaxies residing in the cluster, thus stealing the galaxies from the cluster as they leave.
Filled red triangles show thief groups that have captured a large number (seven or eight) of well-resolved galaxies from the cluster.
Both panels in Figure~\ref{kompaktnost} show that the thief groups have higher values of the compactness parameter $\langle d_{\mathrm{w}} \rangle$, so the thief groups appear to be less compact compared to the field groups in the same mass range. 
As a result of galaxy accretion, thief groups generally contain more members than field groups,
as shown in the right panel of Figure~\ref{kompaktnost}.
On the other hand, non-thief groups seem to have the same properties as the field groups, which makes them indistinguishable.

To further compare the distributions of the field and the cluster groups represented in Figure~\ref{kompaktnost}, we employ statistical tools to test the equality of two two-dimensional samples. Those statistical tests are described in Section~\ref{stat_test_metod}. 
Table \ref{stat_test} shows the p-values for the Cram\'er test, the npdeneqtest, and the Peacock test.
In the first column, we test whether the field and the cluster groups come from the same distribution, as shown in Figure~\ref{kompaktnost}. The low p-values for all the statistical tests used indicate that the field and cluster groups do not follow the same distribution. 
Further, we split cluster groups into thief and non-thief groups and perform the statistics on those sub-samples.
Comparison of field and thief groups results in even lower p-values, which confirms that thief groups occupy different regions in Figure~\ref{kompaktnost}. 
On the other hand, statistical tests for the field and non-thief groups give high p-values, indicating that they follow the same distribution.

A comparison between thief and non-thief groups remains inconclusive.
While low p-values from the Cramér test confirm two distinct samples, the npdeneqtest and Peacock tests yield higher p-values ($p \ge 0.05$), which is close to the commonly used threshold for statistical significance.
The employed tests confirm that groups that go through a galaxy cluster and capture new galaxies from the cluster are less compact and, on average, have more members than other groups in the same mass range. 
However, the cluster passage itself does not influence the compactness of the group. Groups that enter a galaxy cluster as already virialized systems and do not accrete new galaxies follow the same distribution as the field groups in Figure~\ref{kompaktnost}.

\subsection{Thief groups}

 \begin{figure*}[hbt!]
        \centering
        \begin{subfigure}[b]{1\textwidth}
            \centering
            \includegraphics[width=0.3\textwidth]{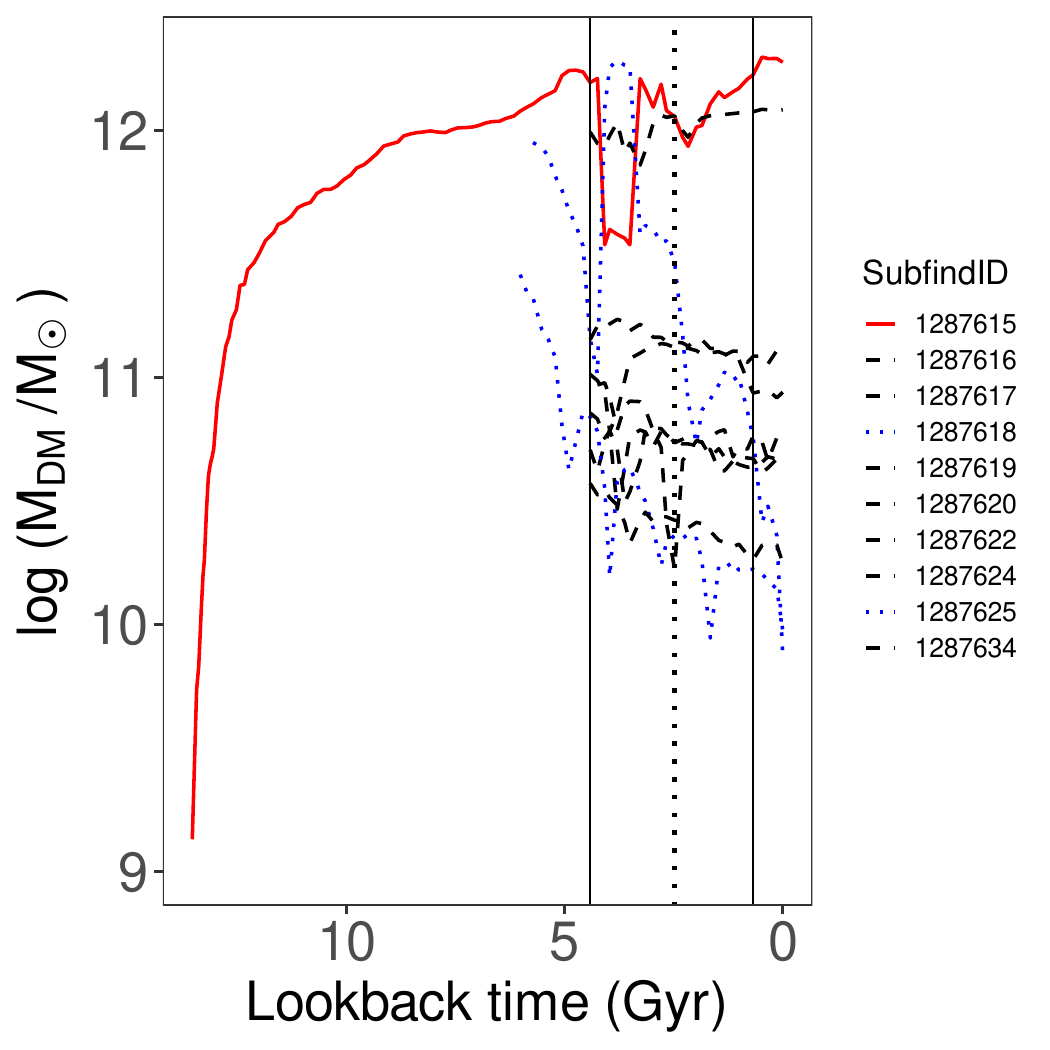} 
          \includegraphics[width=0.3\textwidth]{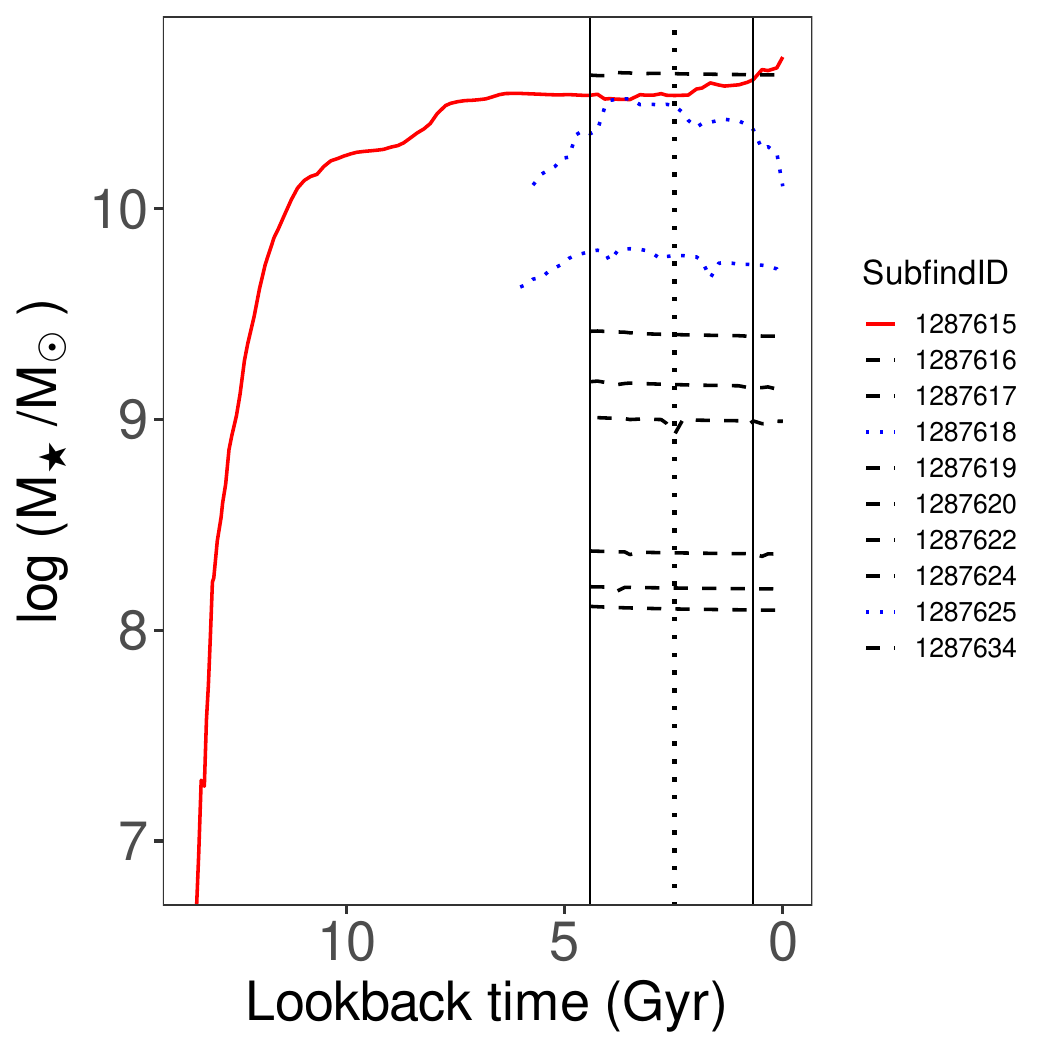} 
            \includegraphics[width=0.3\textwidth]{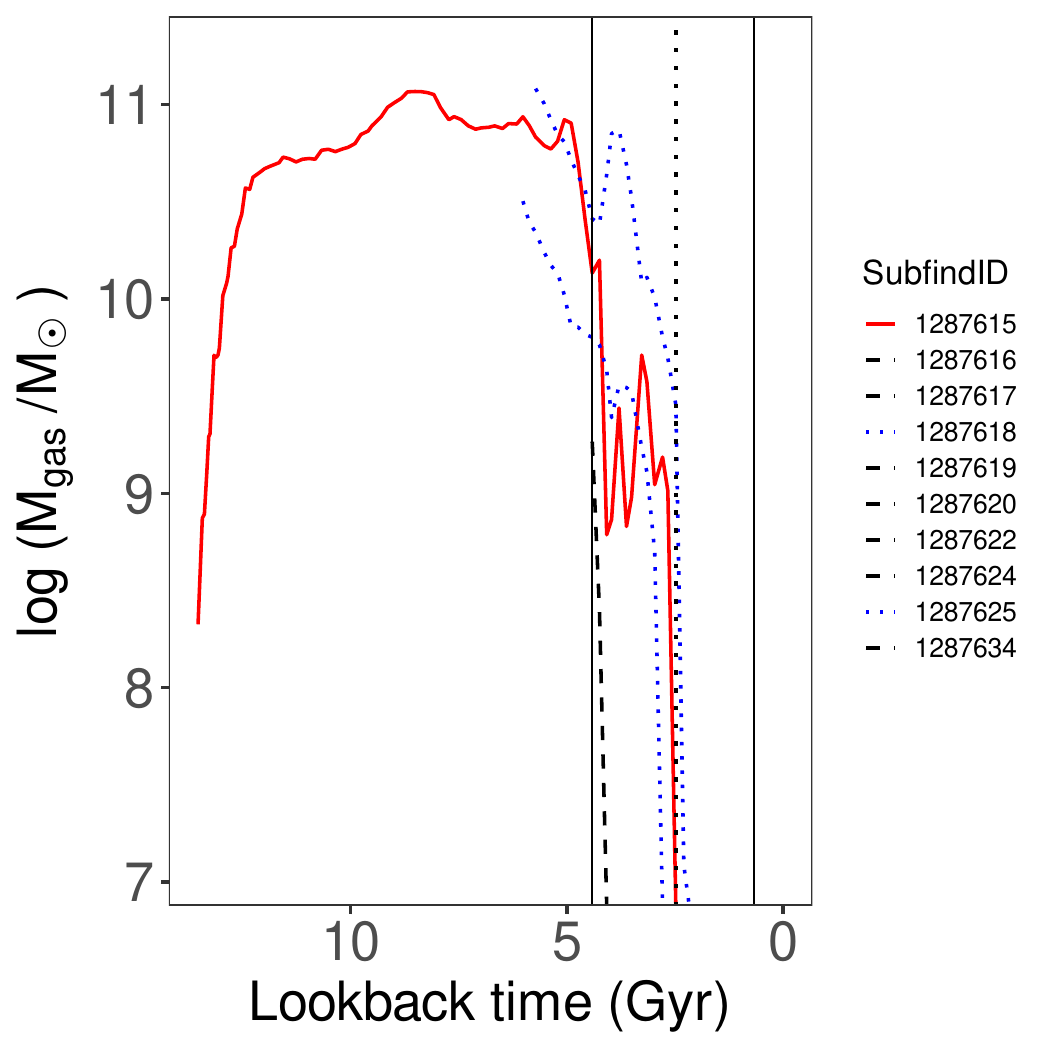} 

            \caption[]%
            {{\small Group ID6627}}  
            \label{ID6627_masa}
        \end{subfigure}
        \hfill
        \begin{subfigure}[b]{1\textwidth}   
            \centering 
            \includegraphics[width=0.3\textwidth]{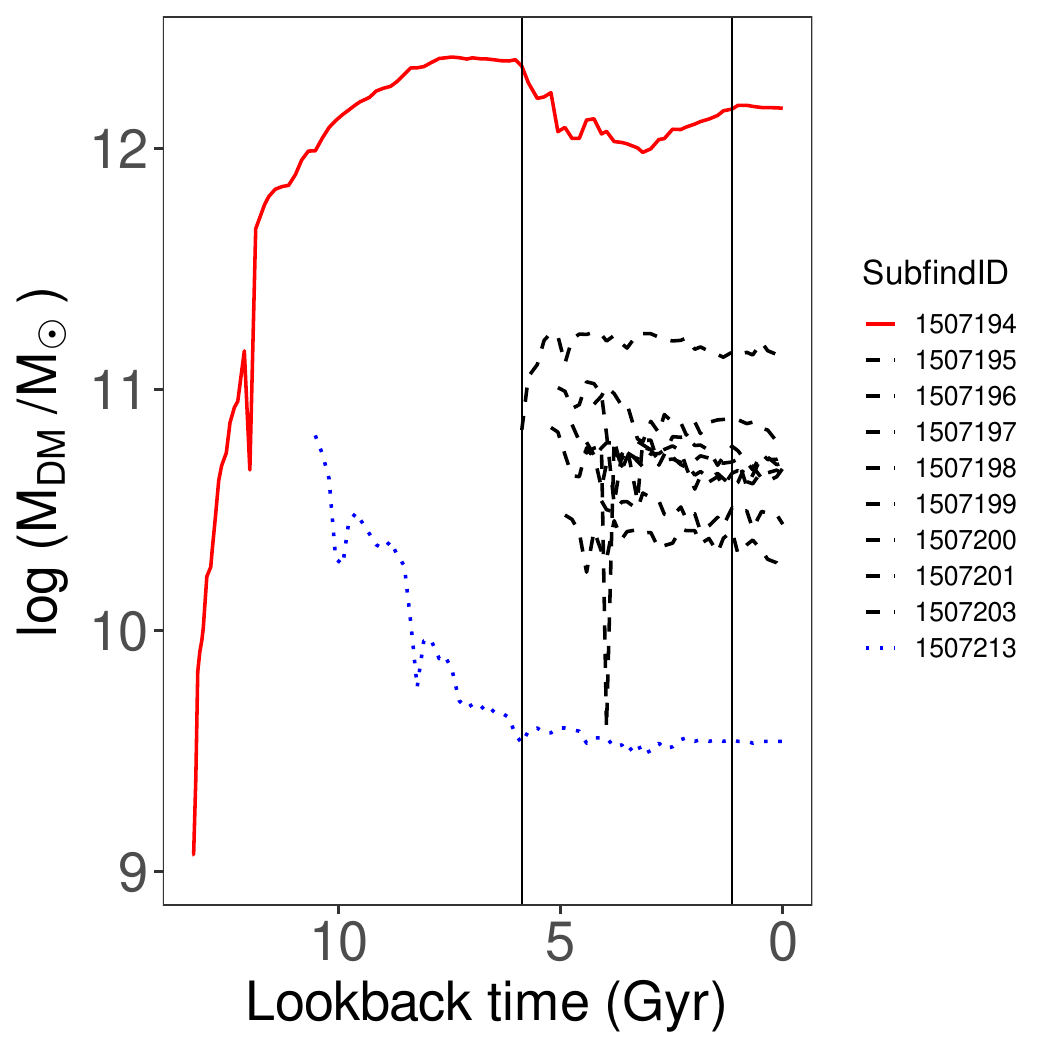} 
          \includegraphics[width=0.3\textwidth]{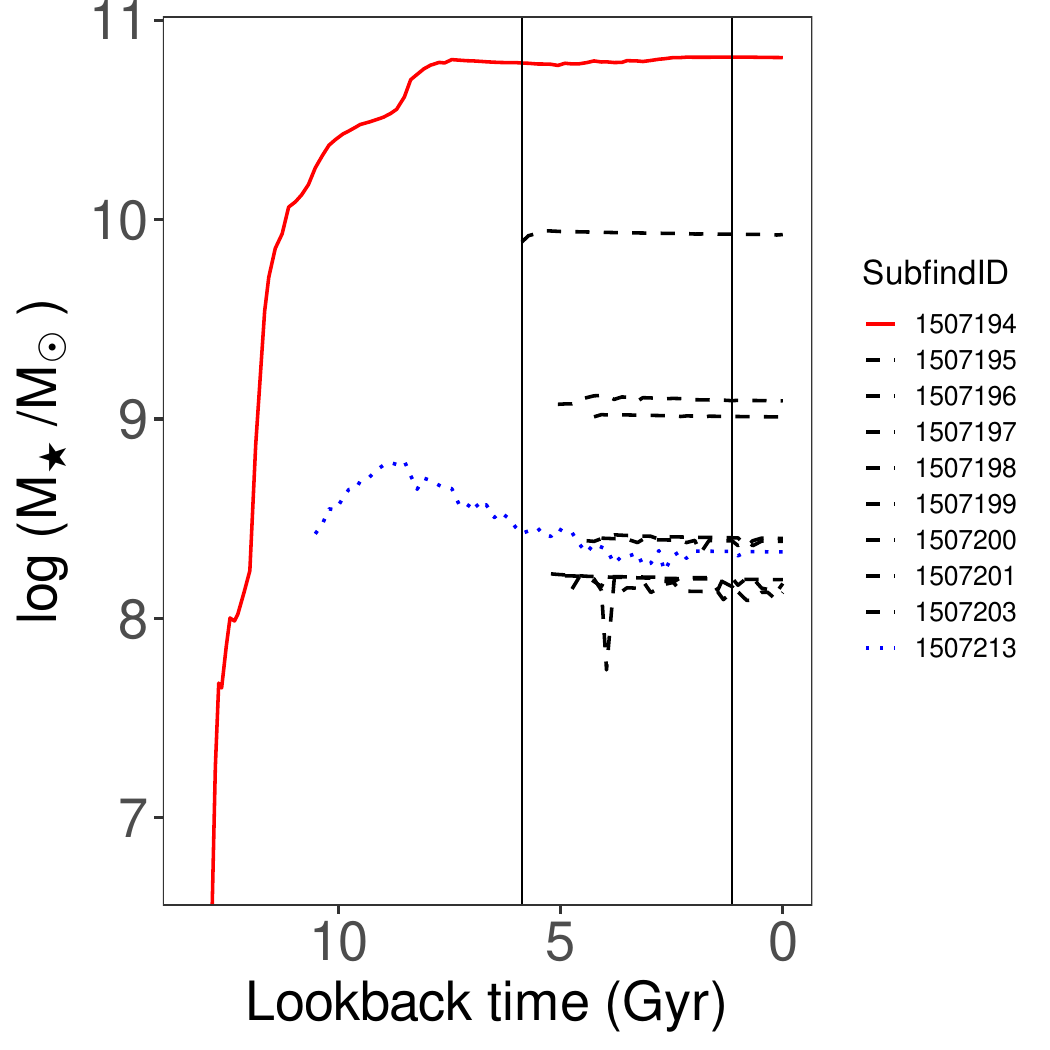} 
            \includegraphics[width=0.3\textwidth]{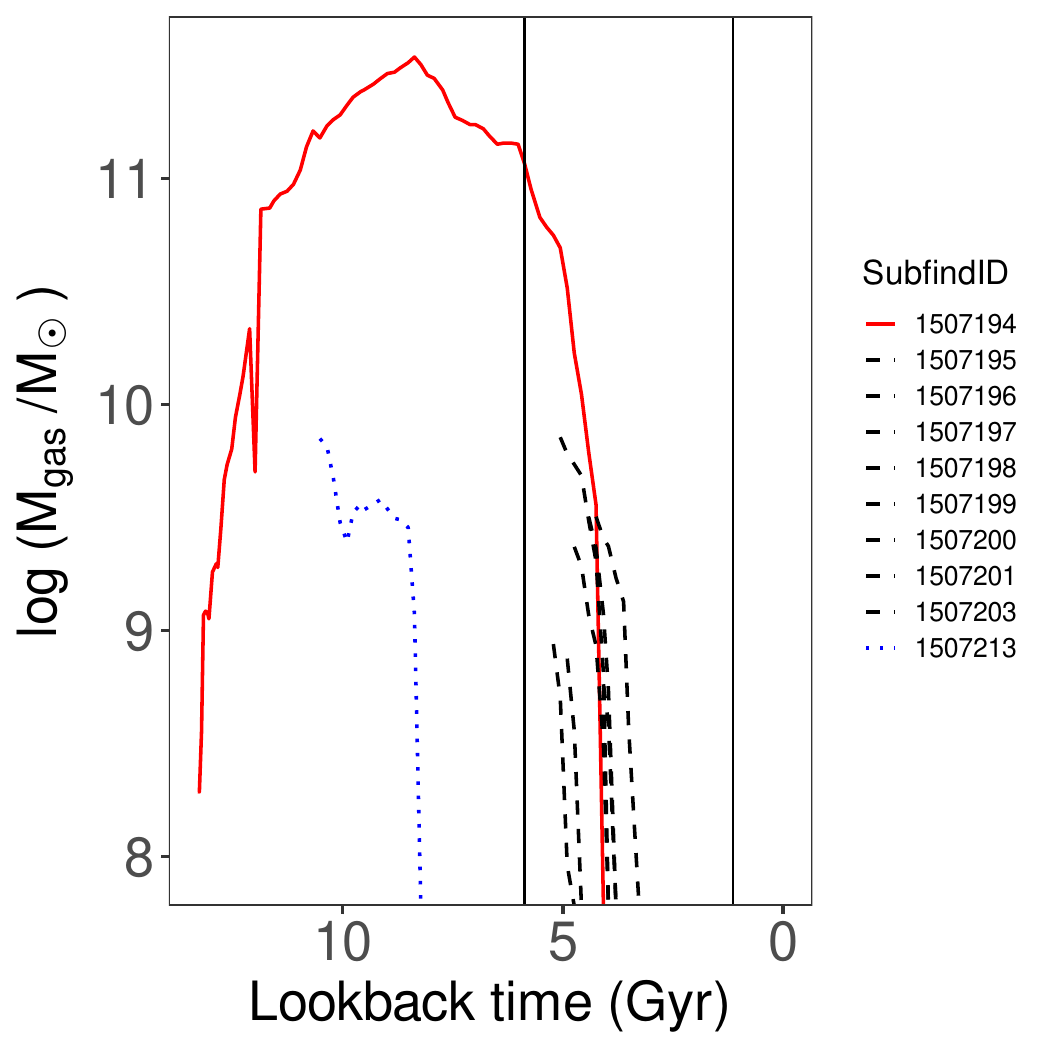} 
            \caption[]%
            {{\small Group ID12982}}  
            \label{ID12982_masa}
        \end{subfigure}
        \hfill
        \begin{subfigure}[b]{1\textwidth}   
            \centering 
            \includegraphics[width=0.3\textwidth]{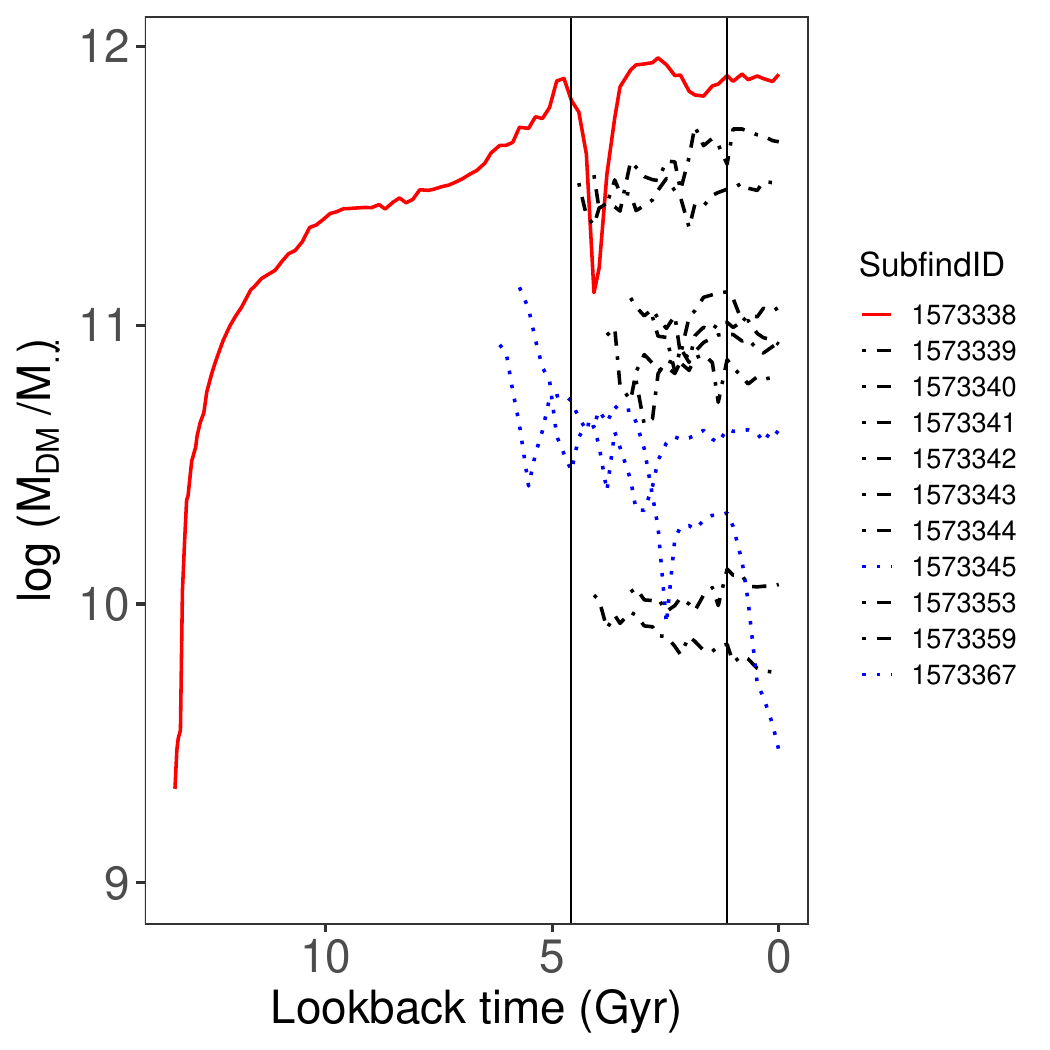} 
          \includegraphics[width=0.3\textwidth]{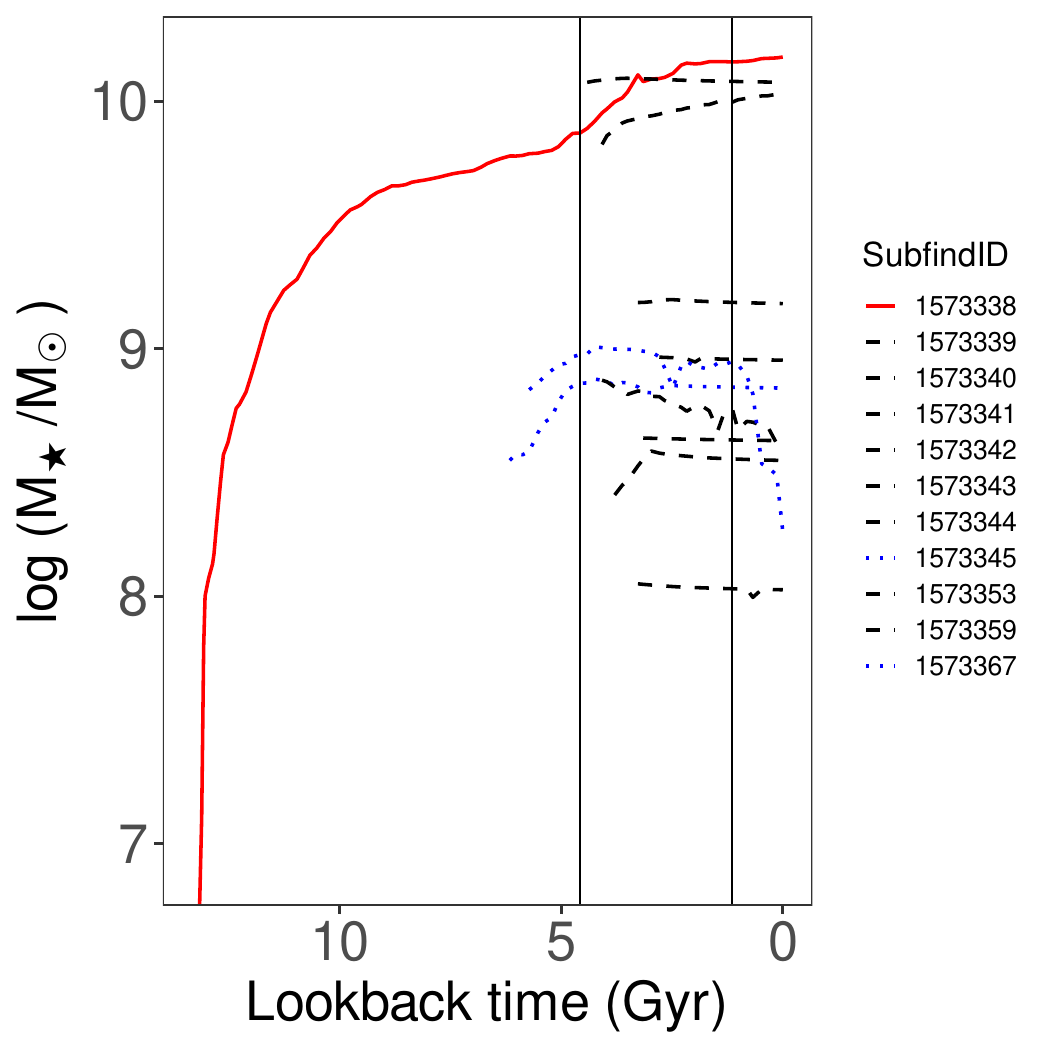} 
            \includegraphics[width=0.3\textwidth]{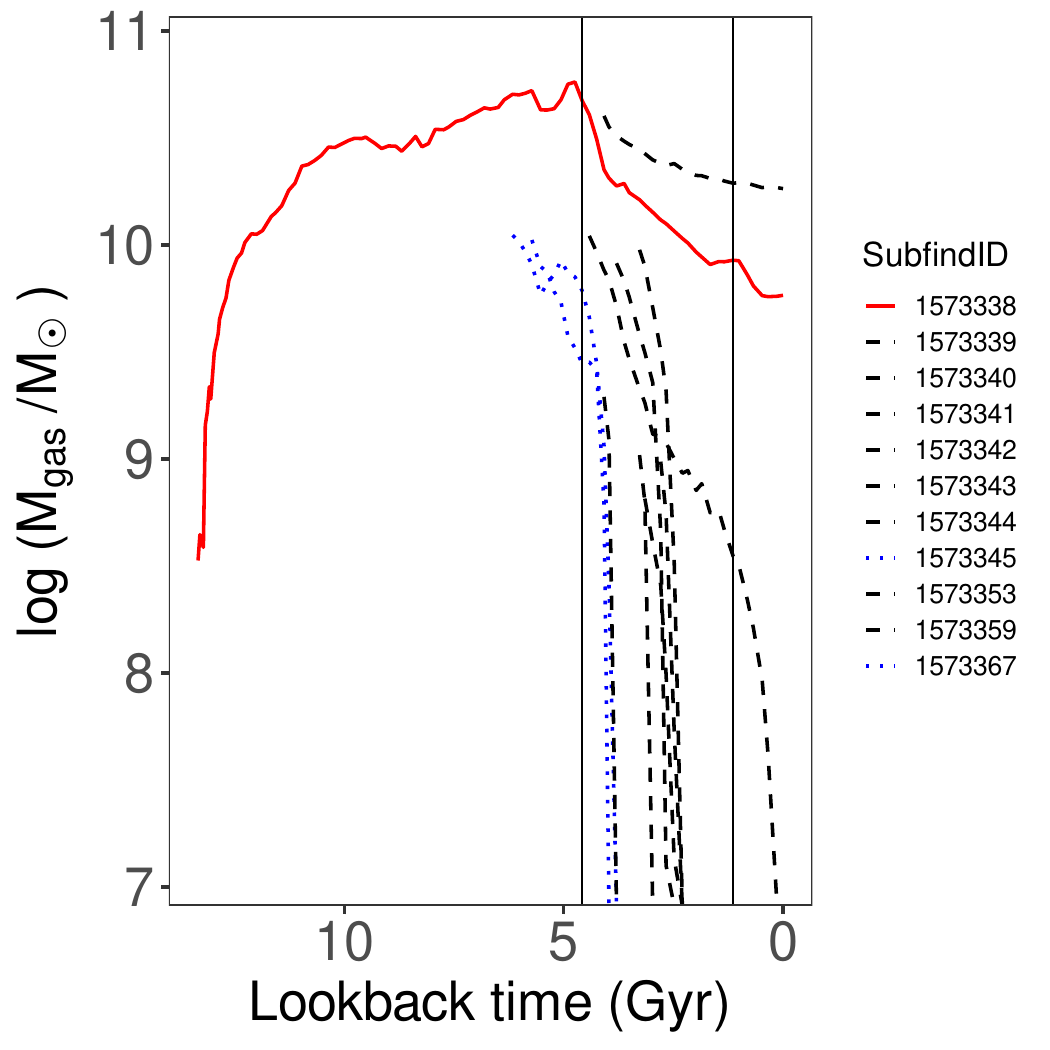}                               
            \caption[]%
            {{\small Group ID15878}}  
            \label{ID15878_masa}
        \end{subfigure}
        \caption{Total DM, stellar and gas mass of group members as a function of the lookback time. Solid vertical lines 
        show the beginning and the end of the cluster phase.
        Red solid lines represent the main galaxy in the group. Blue dotted lines represent 
galaxies that were part of the group before the cluster phase, and black dashed lines represent galaxies
captured during the cluster phase.      
        }
\label{star_mass}
    \end{figure*}
 
In this Section, we explore the unusual evolutionary path of thief groups that enter and then escape a galaxy cluster, capturing a large number of well-resolved cluster galaxies. This sub-sample of the thief groups is marked with filled red triangles in Figure~\ref{kompaktnost}. 
At redshift $z=0$, these groups belong to FoF halos ID6627, ID12982, and ID15878.
The evolution of those groups can be divided into three phases: pre-cluster phase, cluster phase,
and post-cluster phase (after separating from the cluster FoF host halo). 

Figure~\ref{star_mass} shows the total DM, stellar, and gas mass for each group member above the mass
resolution limit as a function of the lookback time.
The red solid lines represent the main galaxy (i.e., the most massive galaxy in the group). The blue dotted lines represent galaxies that were part of the group before the cluster phase, and the black dashed lines represent galaxies captured by the group during the cluster phase. 
The evolution of each satellite galaxy is shown starting from the lookback time at which
the galaxy becomes part of the same FoF halo as the main galaxy, represented by the red solid line.
Solid vertical lines show the beginning and the end of the cluster phase.
These examples show groups of two or three galaxies that make a passage through a cluster and accrete $\geq$~7 galaxies from the cluster.

Figure~\ref{ID6627_masa} represents a special case in which the group experiences a cluster merger during the cluster phase.
At $t_\textrm{lb}=4.41$ Gyr ($z=0.4$) the group merges with a massive cluster with $M=7.05\times 10^{14} \Msun$ and 4741 members. After $\sim2$ Gyr, at $t_\textrm{lb}=2.48$ Gyr ($z=0.2$), this cluster merges with the most massive cluster in the TNG300 simulation, with $M=1.22\times 10^{15} \Msun$ and more than 12000 members. This moment of cluster merger is represented with a dotted vertical line in Figure~\ref{ID6627_masa}. 

\begin{figure*}[hbt!]
        \centering
        \begin{subfigure}[b]{1\textwidth}
            \centering
            \includegraphics[width=0.33\textwidth]{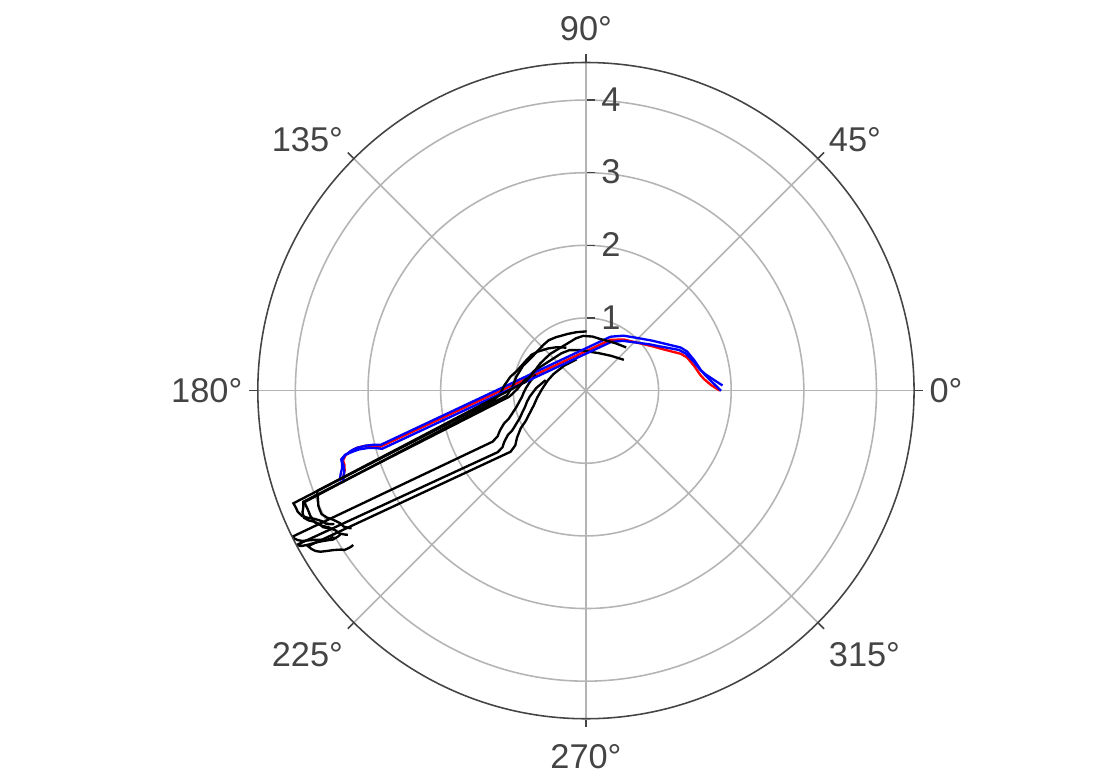} 
            \includegraphics[width=0.33\textwidth]{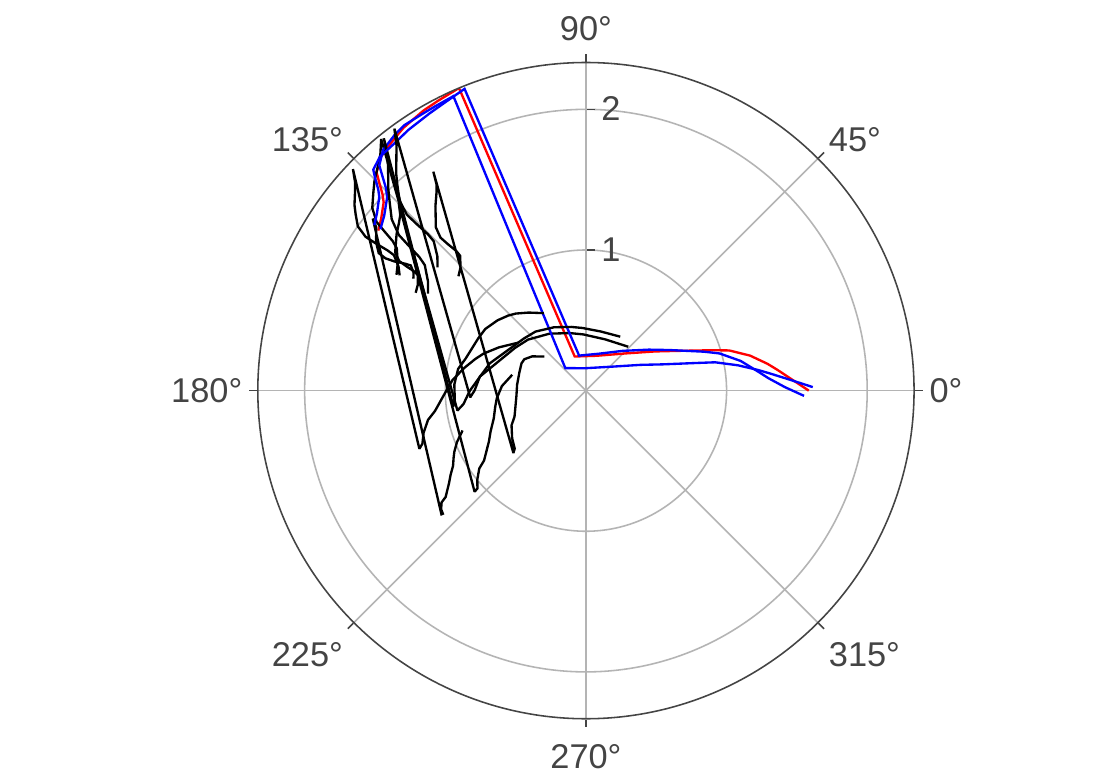} 
            \includegraphics[width=0.33\textwidth]{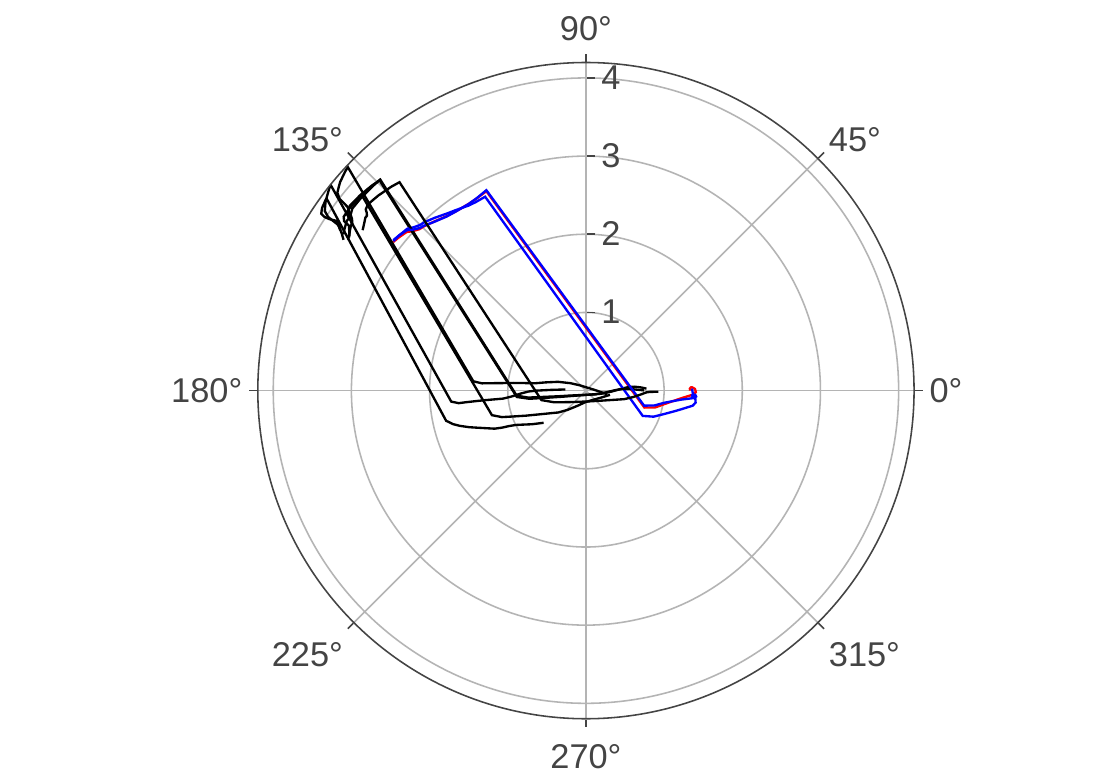} 

            \caption[]%
            {{\small Group ID6627}}  
            \label{ID6627_orbite}
        \end{subfigure}
        \hfill
        \begin{subfigure}[b]{1\textwidth}   
            \centering 
            \includegraphics[width=0.33\textwidth]{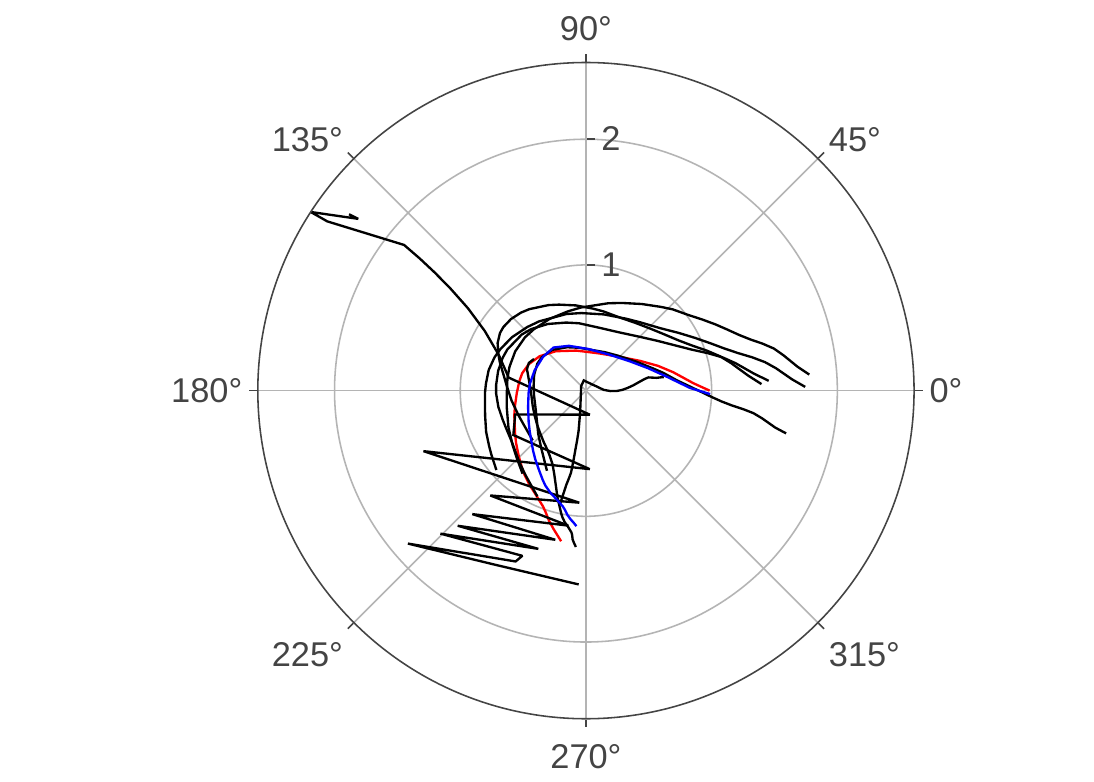} 
            \includegraphics[width=0.33\textwidth]{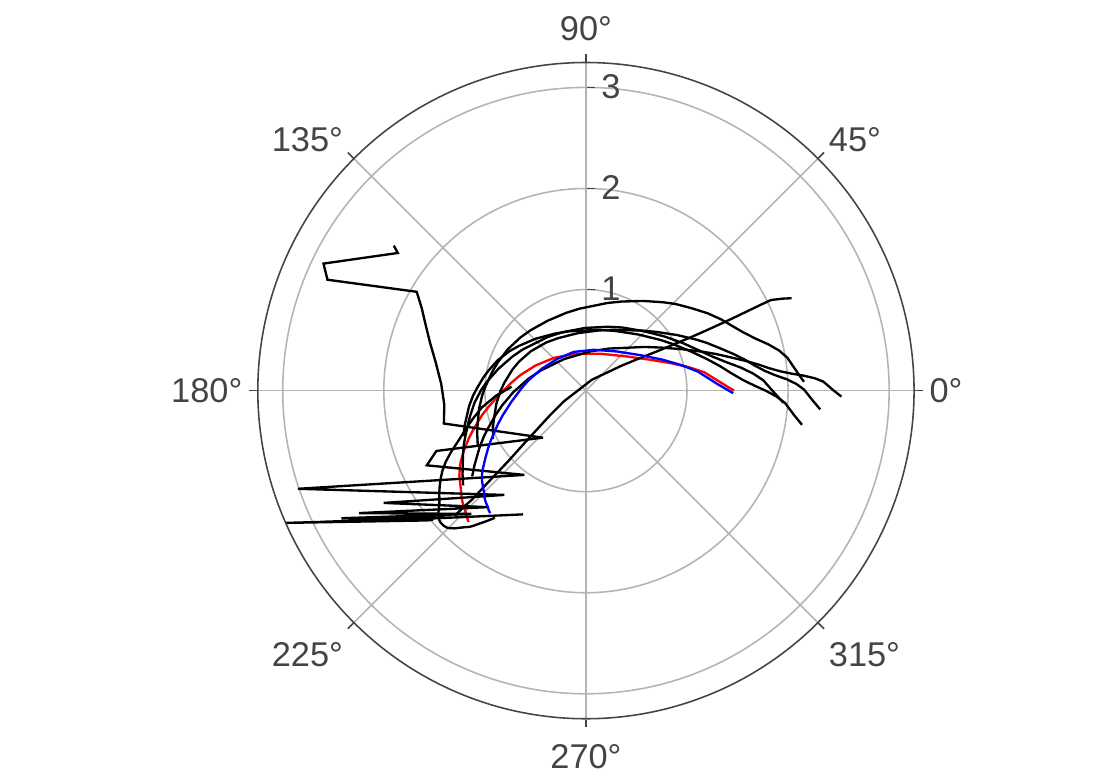} 
            \includegraphics[width=0.33\textwidth]{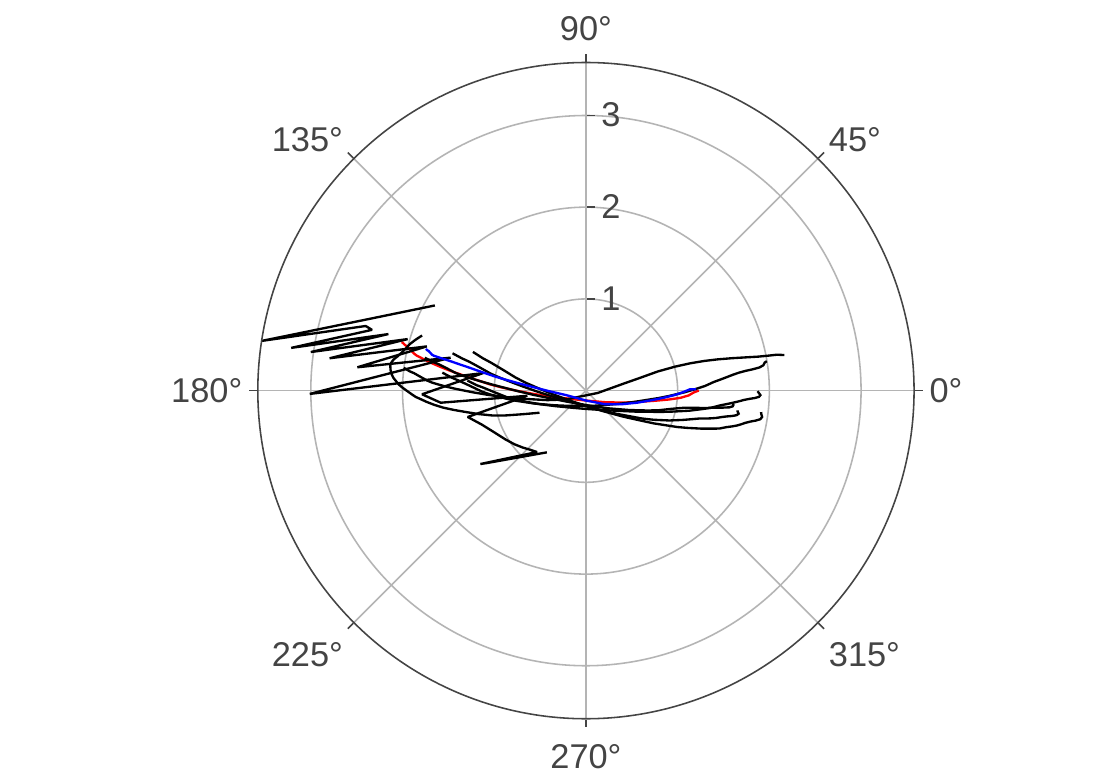} 

                    \caption[]%
            {{\small Group ID12982}}  
            \label{ID12982_orbite}
        \end{subfigure}
        \hfill
        \begin{subfigure}[b]{1\textwidth}   
            \centering 
            \includegraphics[width=0.33\textwidth]{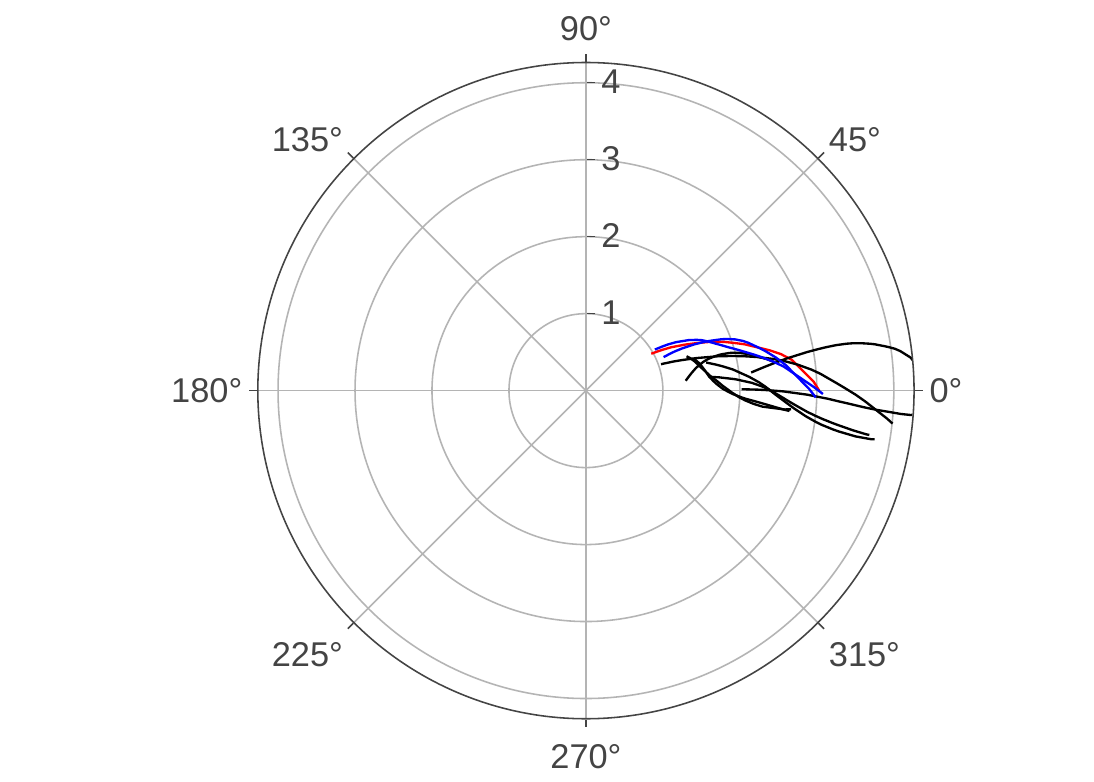}
            \includegraphics[width=0.33\textwidth]{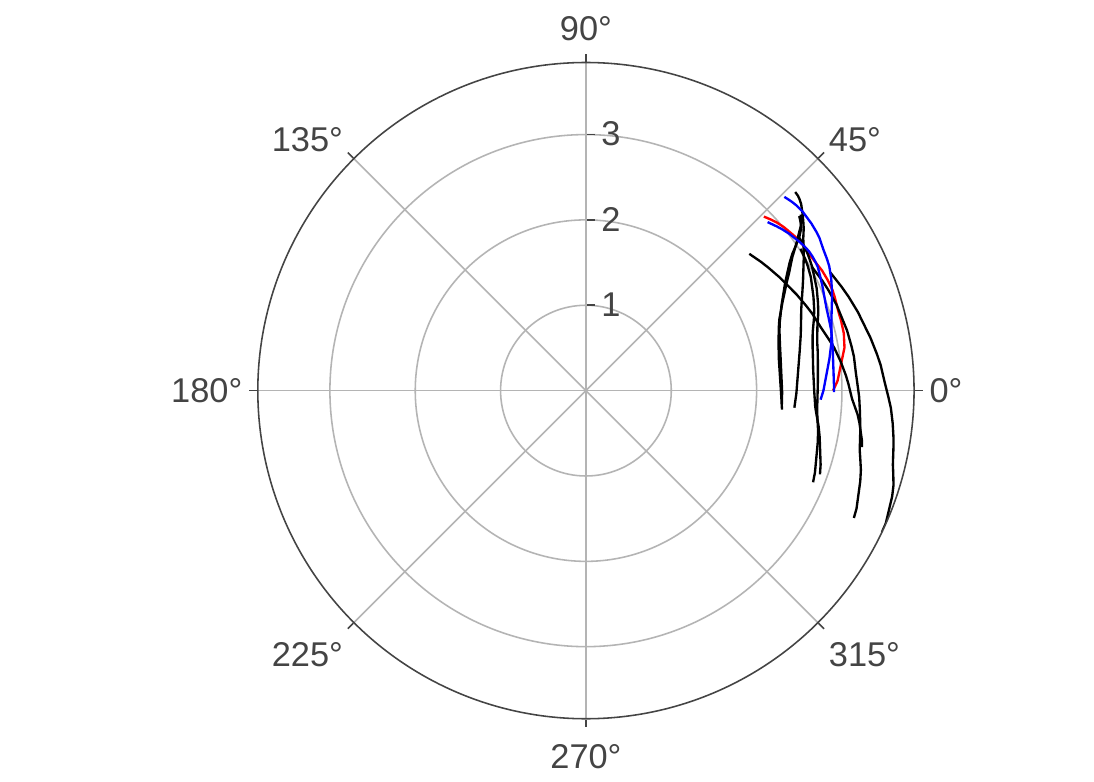}
            \includegraphics[width=0.33\textwidth]{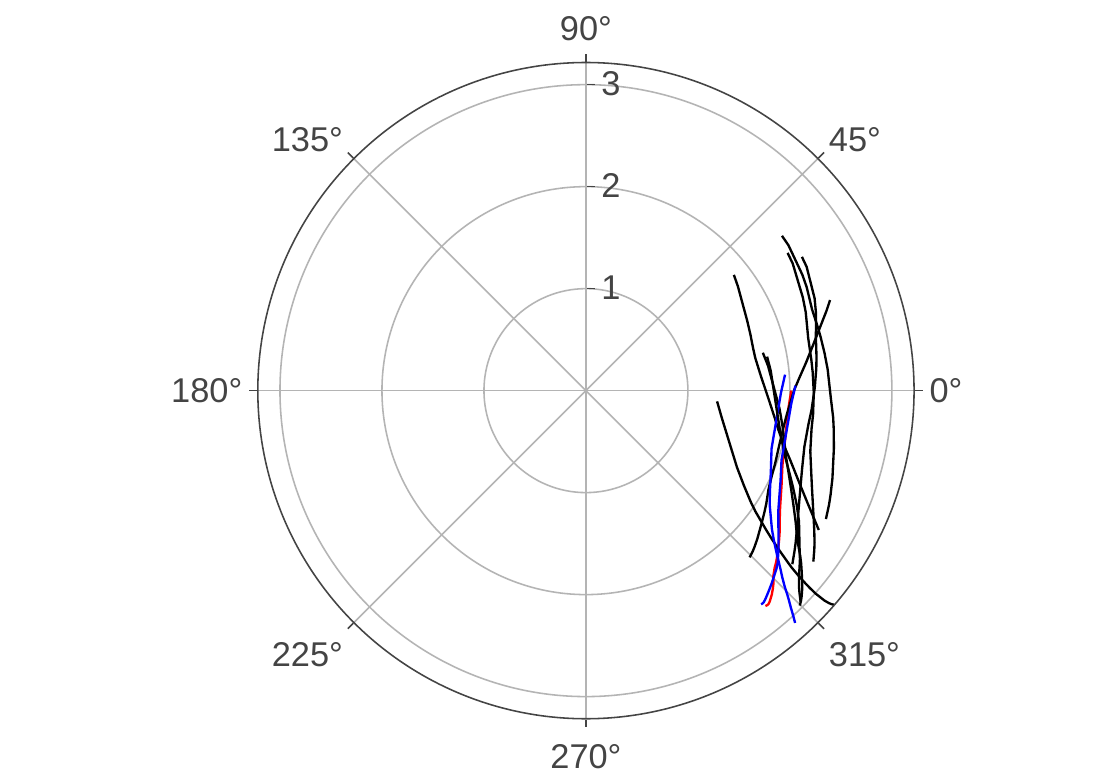}
                                \caption[]%
            {{\small Group ID15878}}  
            \label{ID15878_orbite}
        \end{subfigure}
        \caption{ Orbits of each group member during the cluster phase. Panels represent 'x-y', 'x-z', and 'y-z' projections in polar coordinates, respectively. 
          The same colour notation as in Figure~\ref{star_mass} is used. }
\label{orbite}
    \end{figure*}

Figure~\ref{star_mass} shows that the cluster passage does not influence all
components of the galaxy equally. 
As a group makes a pericentric passage, galaxies experience accretion and tidal stripping
which is mostly limited to DM and gas components at the galaxy outskirts.
The cluster passage shows the greatest influence on the gas component. 
The ID6627 and ID12982 groups have a pericentric passage within the virial radius of the cluster, resulting in gas stripping in all members of the group (Figures~\ref{ID6627_masa} and \ref{ID12982_masa}).
After $\sim2$~Gyr is spent in the cluster environment, the gas is completely removed from all members.
The main galaxy in group ID15878 also shows a decrease in the gas component. However, the gas is still present in the galaxy in the post-cluster phase. In Figure~\ref{orbite}, we show that this group stays at the cluster outskirts and does not experience strong RPS, so the gas is not fully removed. 
During the time spent in the cluster, the galaxy core is protected, and the stellar component does not change significantly.
The increase in stellar mass is noticeable in the main galaxy in group ID15878, while the satellite galaxies show a slight decrease (Figure~\ref{ID15878_masa}).
In all groups, the DM component also shows fluctuations during the cluster phase. 
Figure~\ref{ID6627_masa} shows a sudden and short decrease in the DM component of the main galaxy,
while the satellite galaxy shows the increase at the same snapshot. This is an example of the
'subhalo switching' problem, common during mergers of similar-sized galaxies, where the Subfind 
algorithm incorrectly swaps the identities of the main and satellite galaxies \citep{poole}.

\subsubsection{Orbits of thief groups}

    \begin{figure*}[hbt!]
        \centering
        \begin{subfigure}[b]{0.3\textwidth}
            \centering
            \includegraphics[width=\textwidth]{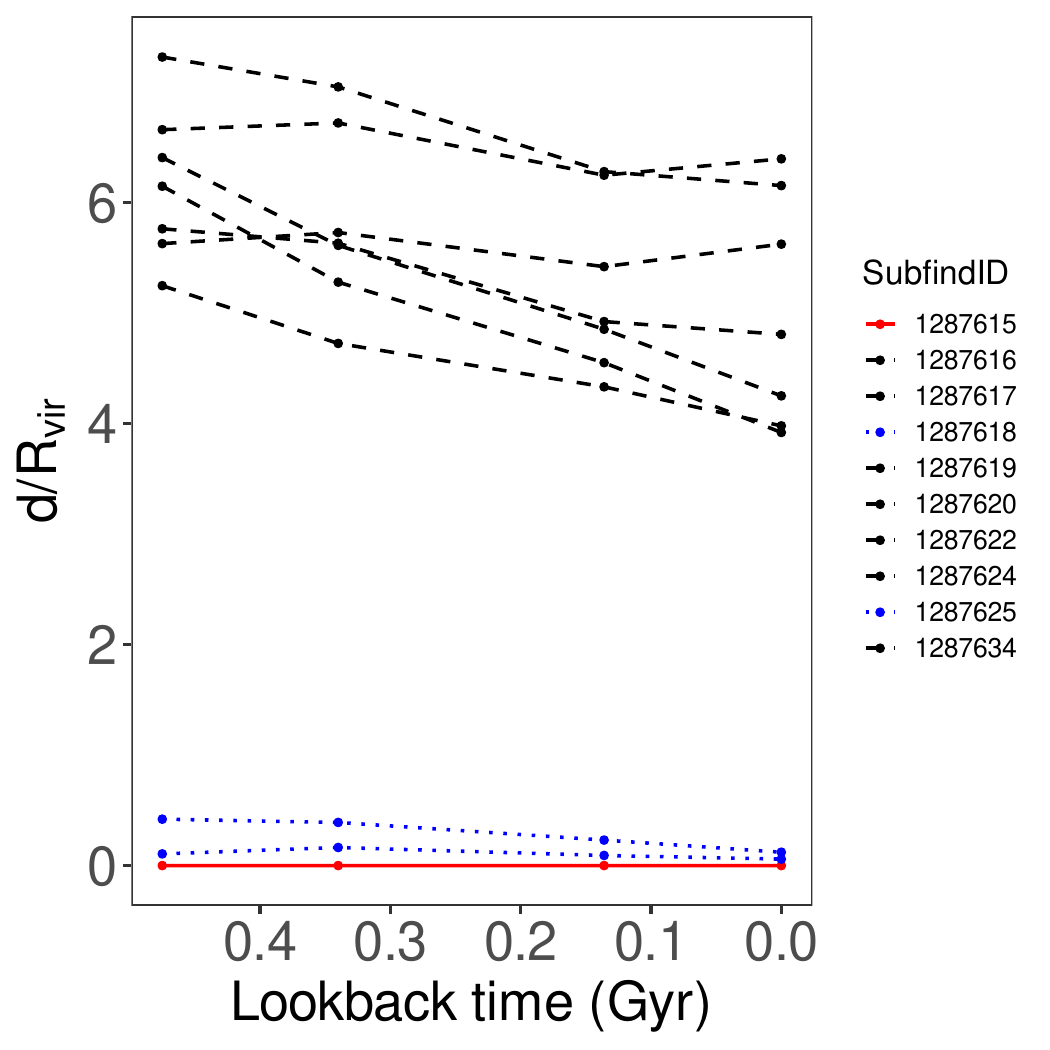} 
            \includegraphics[width=\textwidth]{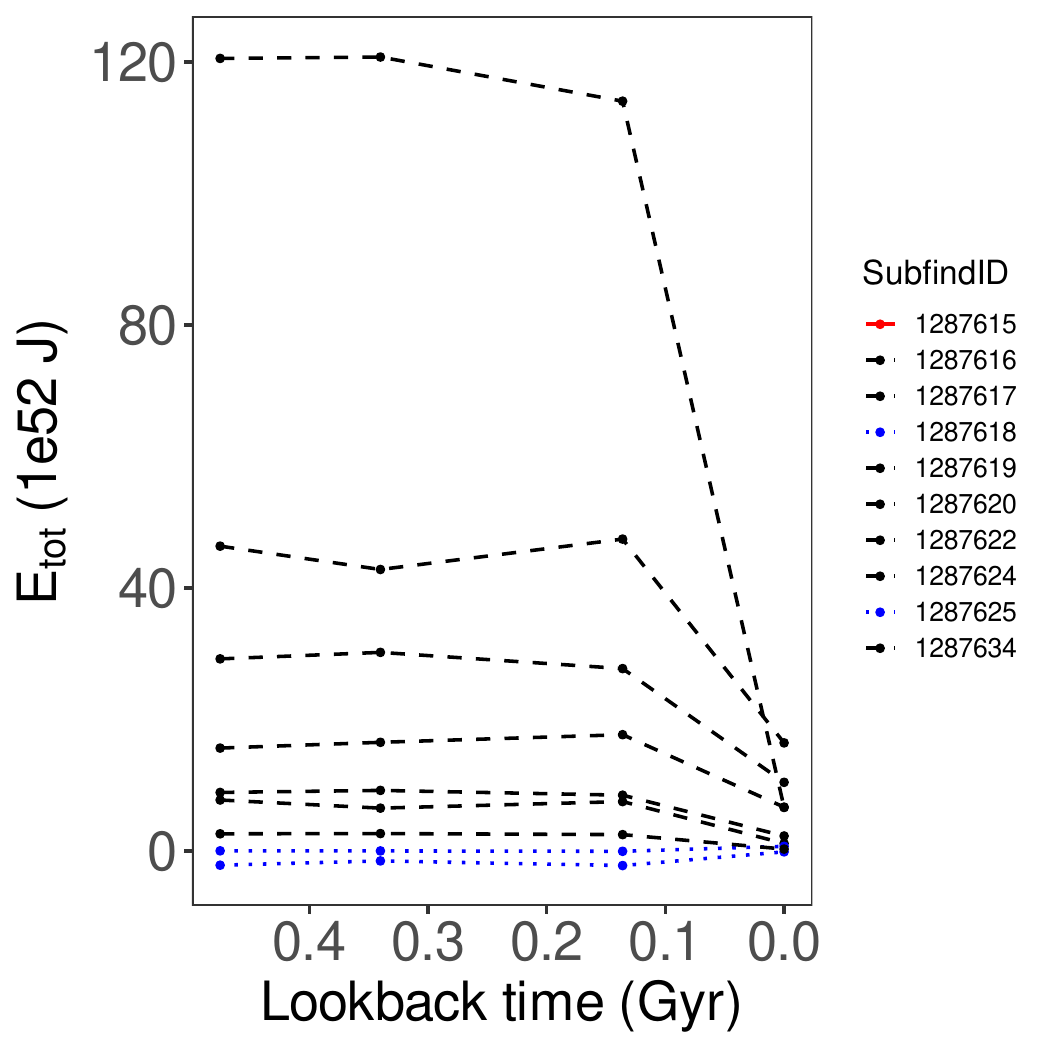} 
            \caption[]%
            {{\small Group ID6627}}  
            \label{ID6627_vezanost}
        \end{subfigure}
        \hfill
        \begin{subfigure}[b]{0.3\textwidth}   
            \centering 
            \includegraphics[width=\textwidth]{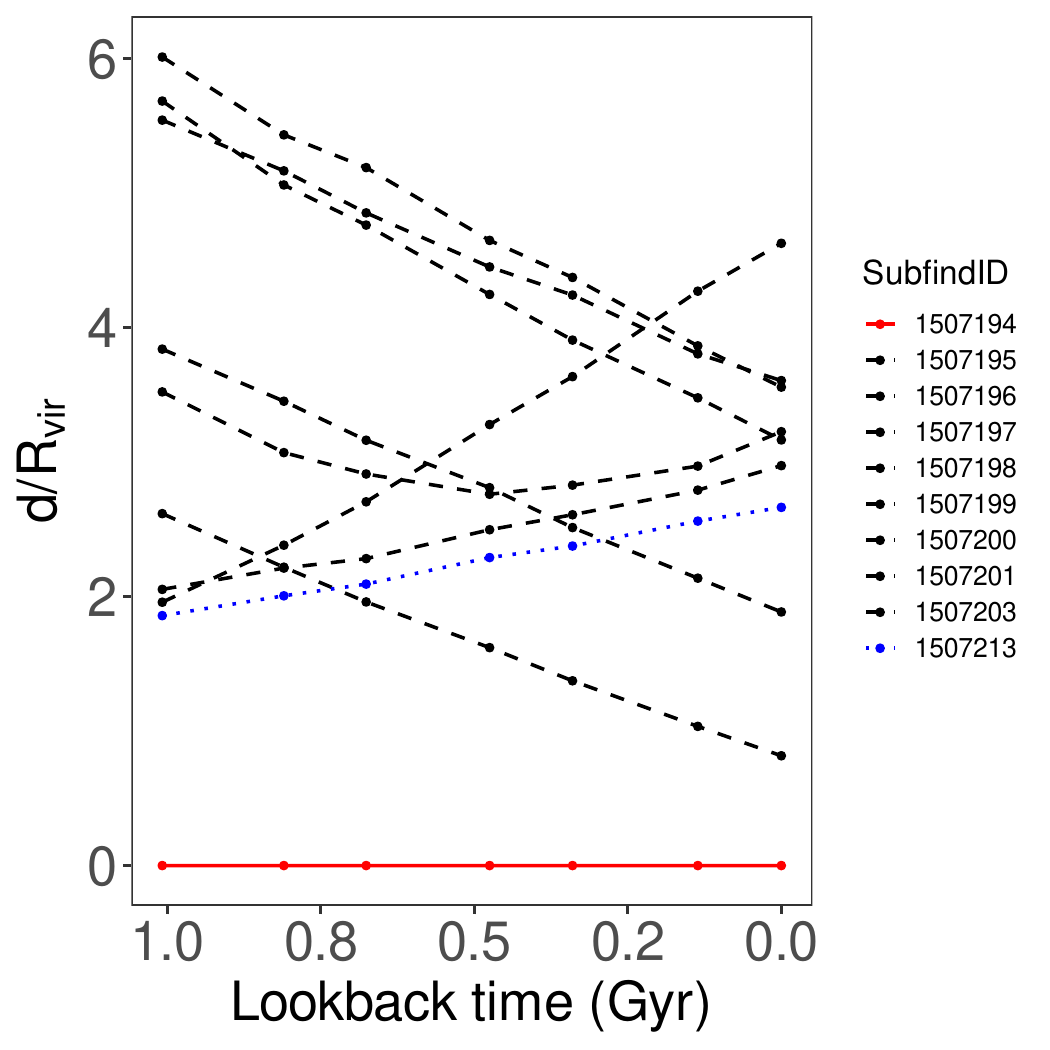} 
            \includegraphics[width=\textwidth]{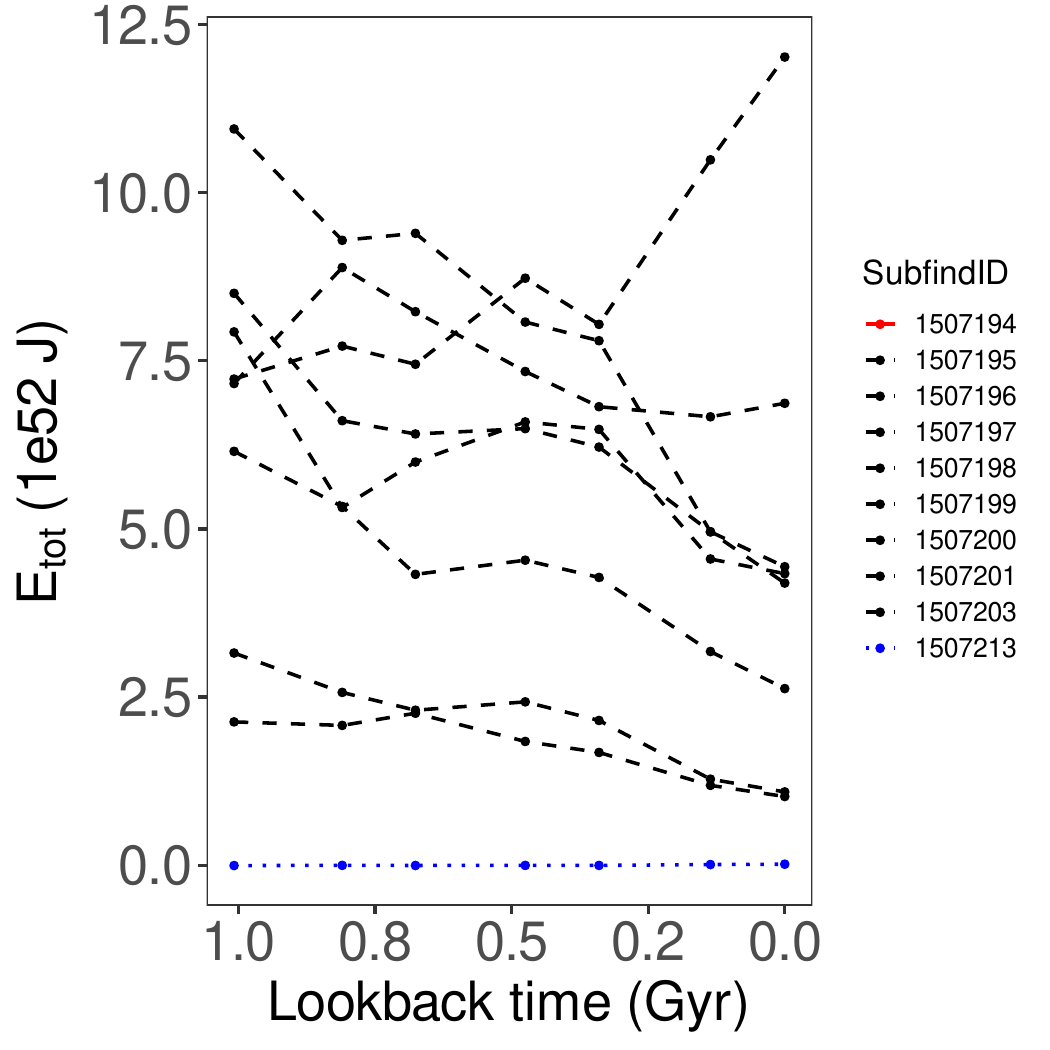} 
                    \caption[]%
            {{\small Group ID12982}}  
            \label{ID12982_vezanost}
        \end{subfigure}
        \hfill
        \begin{subfigure}[b]{0.3\textwidth}   
            \centering 
            \includegraphics[width=\textwidth]{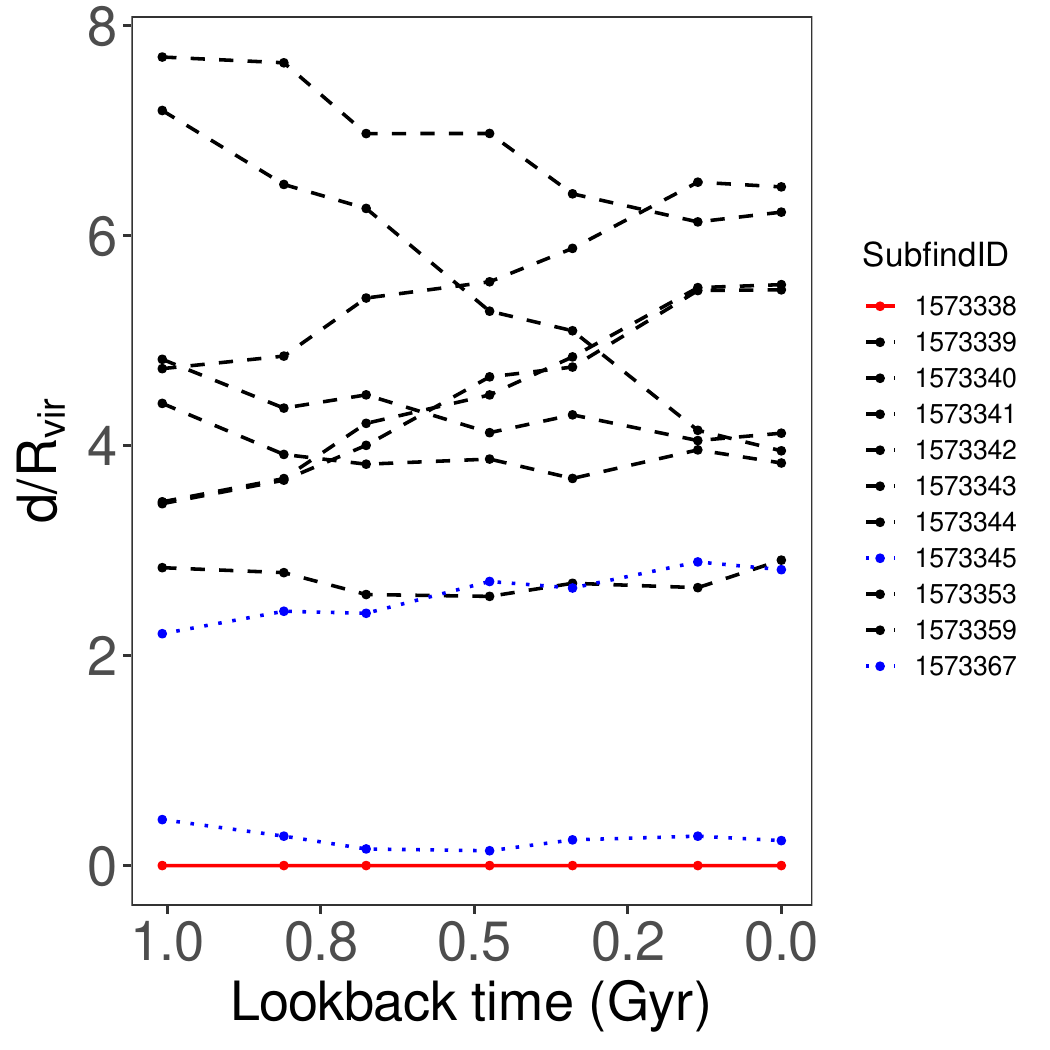}
            \includegraphics[width=\textwidth]{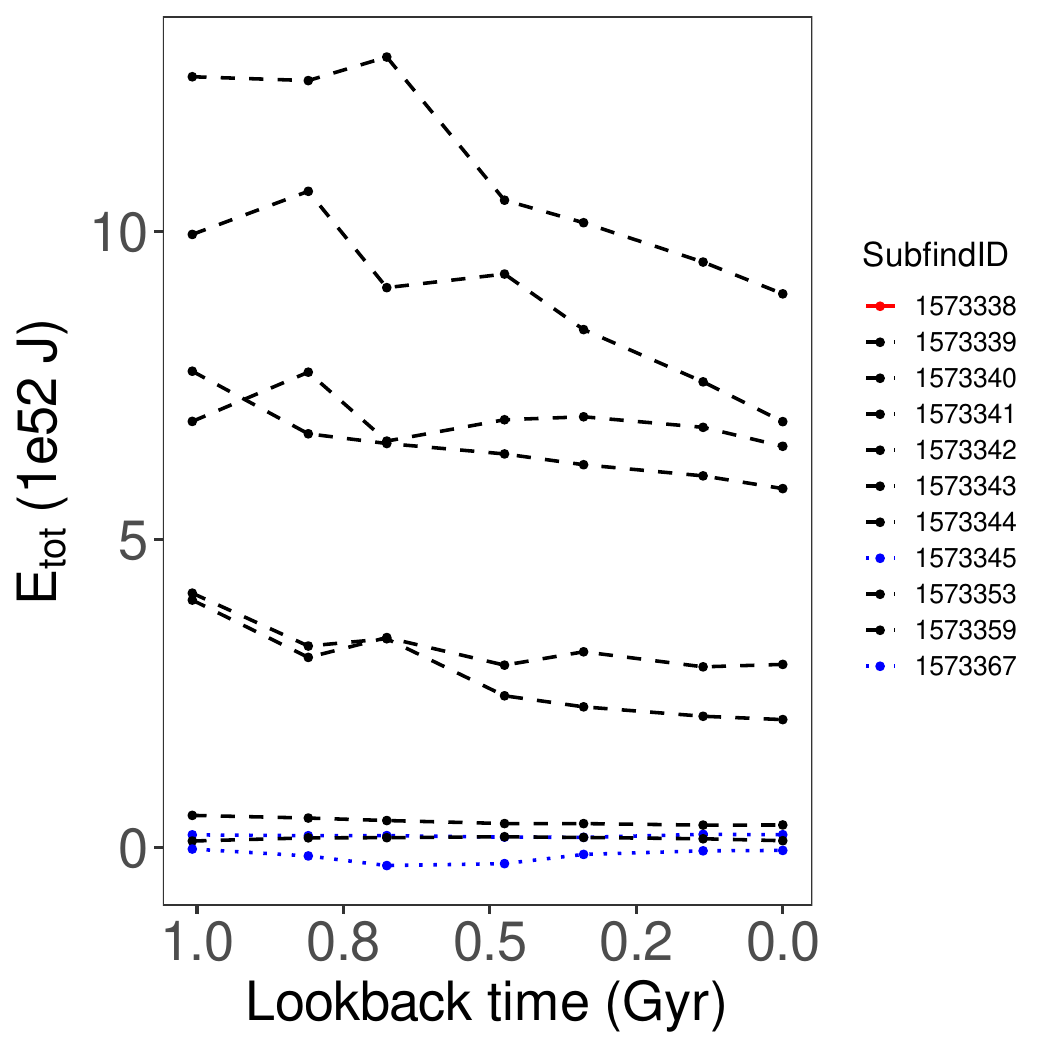} 
                                \caption[]%
            {{\small Group ID15878}}  
            \label{ID15878_vezanost}
        \end{subfigure}
        \caption{ Relative distance (upper panels) and the total energy (lower panels) of group members as a function of lookback time.
          The same notation as in Figure~\ref{star_mass} is used. }
\label{vezanost}
    \end{figure*}

In Figure~\ref{orbite}, we show the orbits of each galaxy in the thief groups during the cluster phase. 
From left to right, panels represent 'x-y', 'x-z' and 'y-z' projections in polar coordinates, respectively.
The polar coordinate system is centred so that $r=0$ corresponds to the cluster centre of mass normalized by its virial radius, and $\phi=0$ is the initial infall angle of the group's main galaxy at the beginning of the cluster phase. 
All three groups perform only one pericentric passage before leaving the cluster.
The ID6627 and ID12982 groups pass through the cluster at distances less than the virial radius of the cluster, while the ID15878 group remains in the outskirts, at $r/R_{\textrm{vir}}\sim3$.
Even though the group that only skimmed the outskirts of the cluster does not experience as strong influence of tidal stripping and dynamical friction as groups that dip
into the cluster's core, both types of interaction can lead to the accretion of new galaxies from the cluster. The potential effects of cluster passage on galaxy morphology, as well as the distinctions between these two scenarios, are further discussed in Section \ref{morphology}.

Figure~\ref{ID6627_orbite} shows that group ID6627 resides inside the virial radius of the cluster at
the moment when it merges with the most massive cluster in the simulation. 
The group stays on the outskirts of this massive cluster, at $r/R_{\textrm{vir}}>3$, until it leaves the cluster halo at $t_\textrm{lb}=0.47$ Gyr ($z=0.03$). The first panel in Figure~\ref{ID6627_orbite} shows that the accreted galaxies, represented with black lines, belong to two different groups of galaxies.
Figure~\ref{ID12982_orbite} shows that group ID12982 has a relatively close pericentric passage to the centre of the cluster, at $r/R_{\textrm{vir}}<0.5$. Most of the accreted galaxies have orbits similar to those of the original group members (represented in red and blue), with the exception of one
accreted galaxy with an unusually eccentric orbit.

In Figure~\ref{vezanost}, we show the evolution of these three groups in the post-cluster phase,
thus after escaping the cluster. 
The upper panels of Figure~\ref{vezanost} show the relative distance of each group member from the centre of mass of the group as a function of lookback time. The distance is normalized by the group's halo virial radius ($R_{\textrm{vir}}$).
To test whether galaxies are gravitationally bound to the group, we compare their kinetic and potential energy. The sum of kinetic and potential energy, thus, the total energy, is negative for gravitationally bound systems.
The lower panels of Figure~\ref{vezanost} show the total energy of each group member as a function of the lookback time. The same notation as in Figure~\ref{star_mass} is used.        
Captured galaxies generally reside at the outskirts of the group, while galaxies that have been part of the group before the cluster passage stay within $\leq2R_{\textrm{vir}}$. 
However, as a group leaves the cluster, the captured members gradually become
more gravitationally bound. This behaviour is well represented in Figure~\ref{ID6627_vezanost}. 
The decrease in total energy is also visible for almost all captured galaxies, with the exception
of two members of the ID12982 group, shown in Figure~\ref{ID12982_vezanost}. 
This group is in the process of ejecting two previously accreted galaxies, as 
a consequence of gravitational interactions within the group.
The general decrease in the total energy of the system, seen in Figure~\ref{vezanost},
suggests that right after the cluster phase the group is still not a virialised system.

\subsubsection{Galaxy compactness in thief groups}

\begin{figure}[hbt!]
\centering
\includegraphics[width=0.99\columnwidth,keepaspectratio=true]{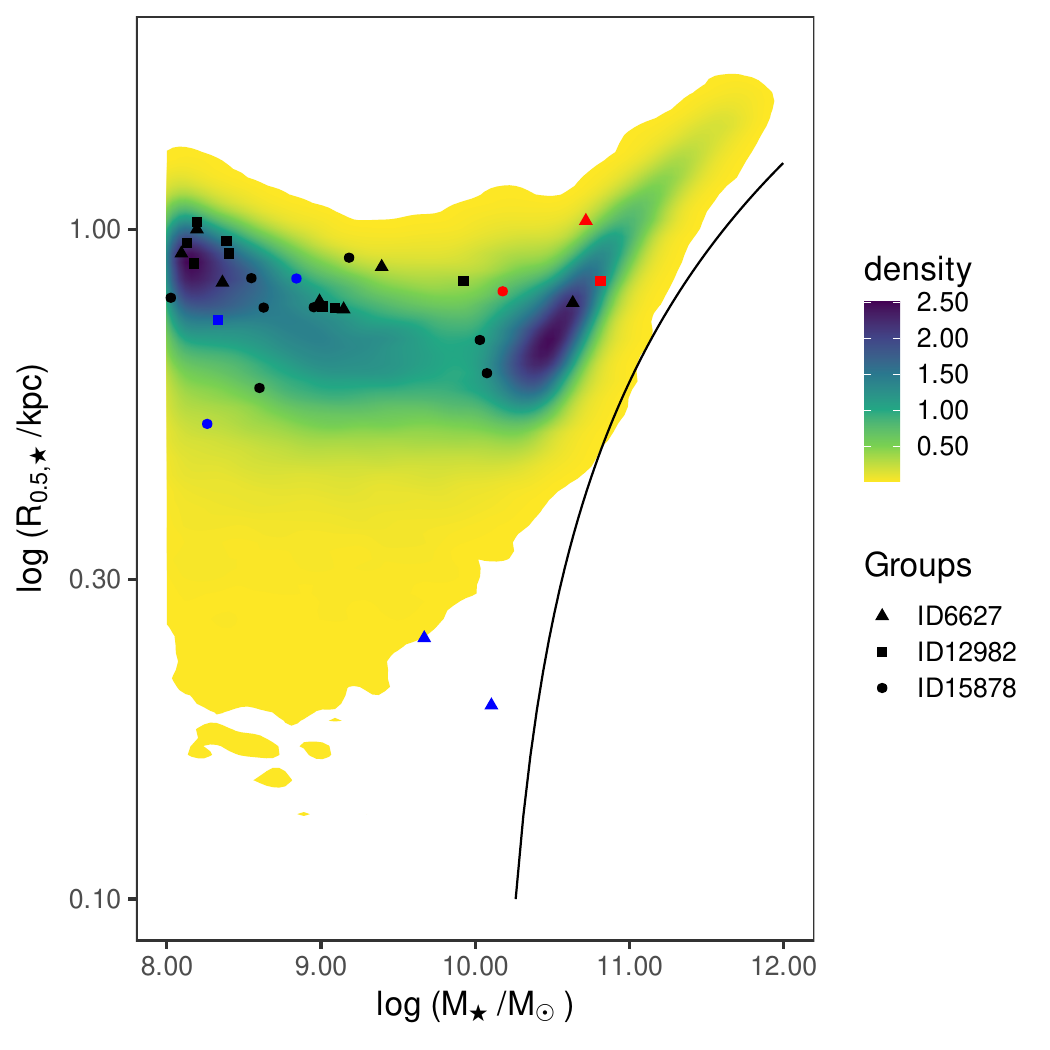}
\caption{colour-coded distribution of field group galaxies at $z=0$ in the mass-size plane.
Different symbols represent galaxies belonging to different thief groups. The same colour symbols 
as in Figure~\ref{star_mass} are used. Solid black line represents compactness
criterion given by Equation~\ref{kriterijum}.
}
\label{r-m}
\end{figure}

\begin{figure}[hbt!]
\centering
\includegraphics[width=0.99\columnwidth,keepaspectratio=true]{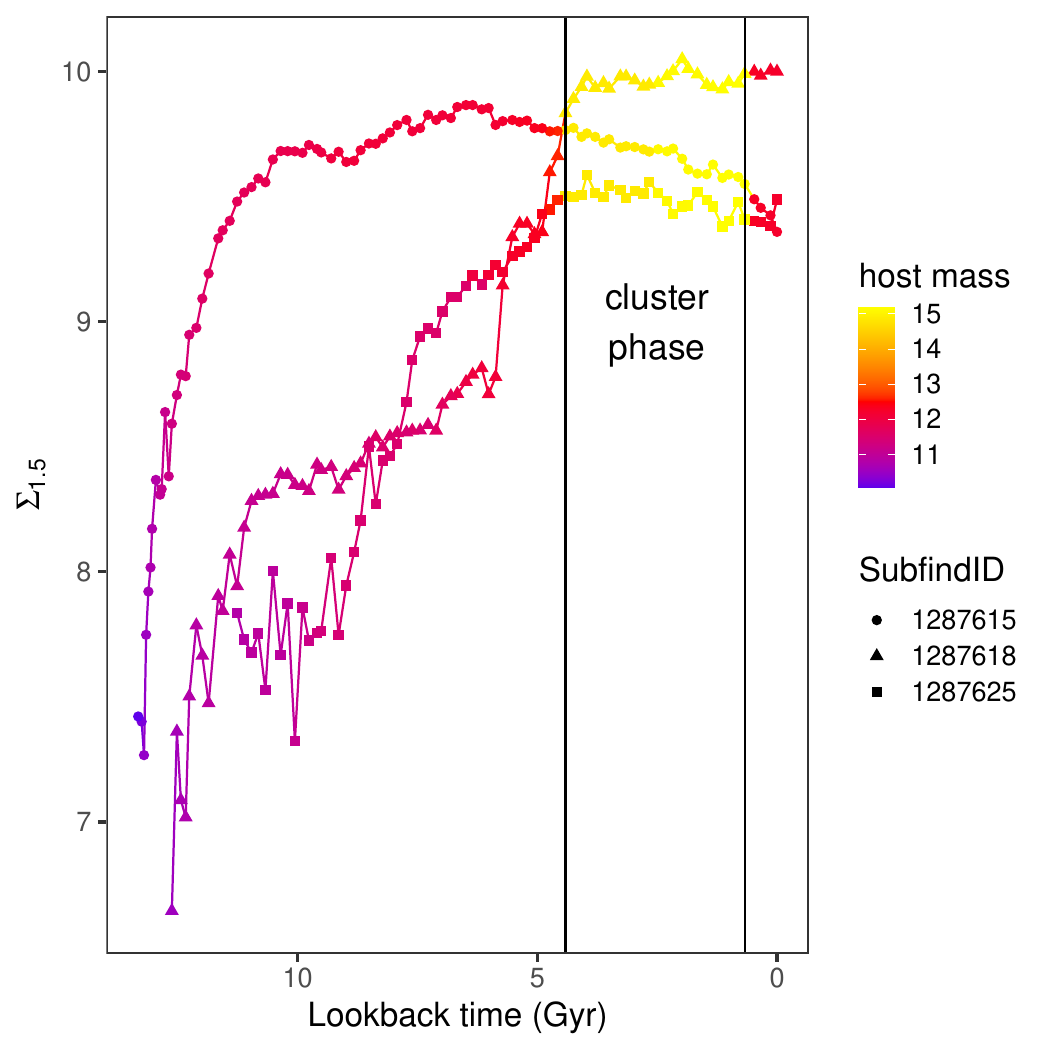}
        \caption{ Evolution of compactness parameter $\Sigma_{1.5}$ in group ID6627
as a function of the lookback time.  Circles represent the main galaxy, while triangles and squares 
represent satellite galaxies. Different colours indicate the total host halo mass in units
$\log M_{\textrm{host}}/\Msun$.}
\label{kompaknost_galaksija}
    \end{figure}

Thief groups represent an example of groups of galaxies that had been processed by a galaxy cluster and managed to escape the cluster environment. 
The evolution of their components, represented in Figure~\ref{star_mass}, shows signs of tidal stripping.
Here, we explore whether the cluster passage influences the compactness of galaxies in thief groups, as suggested by \citet{Chilingarian}.

Galaxy compactness can be quantified using observationally determined criteria, such as
\citet{Barro+2013ApJ...765..104B}:
\begin{equation}
\begin{aligned}
\Sigma_{1.5}\equiv \log \frac{M_\star}{(0.75\cdot R_{0.5,\star})^{1.5}} \big[\frac{\Msun}{\textrm{kpc}^{1.5}}\big] > 10.3.
\end{aligned}
\label{kriterijum}
\end{equation}
\noindent where compact galaxies are defined as those with parameter $\Sigma_{1.5}>10.3$.
Calculating $\Sigma_{1.5}$ for galaxies in thief groups shows that none of the considered galaxies satisfied the compactness criterion. 
The most compact galaxy in our thief sample, SubfindID 1287618, has $\Sigma_{1.5}=10$.
At $z=0$ the galaxy has $M_\textrm{tot}=2.03\times10^{10} \Msun$, $M_\star=1.27\times10^{10} \Msun$, no gas, and DM to baryon ratio $M_\textrm{DM}/M_\textrm{B}=0.59$, with half-mass stellar radius $R_{0.5,\star}=1.56$~kpc.

However, most of the observationally based criteria can be overly restrictive.
An alternative approach, based on the notion that compact galaxies are outliers on the mass-size relation \citep[used by, e.g.,][]{Lohmann+2023MNRAS.524.5266L}, is more suitable, especially considering the fact that the mass-size relation is well-reproduced by IllustrisTNG \citep{Genel+2018MNRAS.474.3976G}.
We calculate the mass-size relation for galaxies in field groups and show the positions
of galaxies from thief groups in that relation. 
The mass-size relation for field galaxies at $z=0$ is shown in Figure~\ref{r-m}. 
The distribution of field galaxies is represented as a density plot,
where the colour bar indicates the estimated 2D density of data points.
Galaxies belonging to different thief groups are shown with different symbols. Similarly to the notation used in Figure~\ref{star_mass}, red symbols represent the main galaxies, the blue symbols
represent the satellite galaxies that entered the cluster together with the main galaxy, and the black symbols represent the galaxies accreted during the cluster phase. 
Solid black line represents observationally determined compactness
criterion given by Equation~\ref{kriterijum}. None of the considered galaxies
occupies the right-hand region of the plot, where compact galaxies are expected to be found.
However, satellite galaxies in group ID6627 represent two outliers from the mass-size relation,
being significantly more compact than field galaxies. 
Further, we briefly describe the evolution of those fairly compact galaxies, with SubfindIDs 1287618 and 1287625 at $z=0$.

The unusual evolution of those galaxies starts in isolation, where they grow mostly by accretion.
At $z=0.6$ ($t_\textrm{lb}\sim 6$ Gyr), during two successive snapshots, both galaxies become part of the same group that resides in the outskirts of a cluster. 
At $z=0.4$ ($t_\textrm{lb}=4.41$ Gyr), the host group merges with a massive cluster
with $M=7.05\times 10^{14} \Msun$. During the passage of the group within the virial radius of the cluster,
the cluster merges with the most massive cluster in the simulation. 
At $t_\textrm{lb}=0.47$ Gyr, the group leaves the cluster environment. During the cluster phase, the group members remained gravitationally bound, and more members from the cluster were accreted. 

While our primary focus was on compact galaxies, we have to acknowledge the fact that environmental effects can lead to the formation of ultra-diffuse galaxies (UDGs) as well \citep[e.g.,][]{Jiang+2019, Sales+2020MNRAS.494.1848S, Tremmel+2020MNRAS.497.2786T, Benavides+2023MNRAS.522.1033B}, the other class of galaxies that can represent outliers on the mass-size relation (i.e., in the upper end). Although a few galaxies in thief groups lie slightly above the average size for their stellar mass (see Figure~\ref{r-m}), none of them reach the size and surface brightness thresholds typically used to define UDGs \citep{vanDokkum+2015ApJ...798L..45V}. Thus, our data do not support the formation of UDGs through the cluster processing scenario explored in this work. This does not rule out cluster processing as a viable pathway for UDG formation; rather, it may indicate that longer times spent in the cluster environment, or different orbital histories, are required.

In Figure~\ref{kompaknost_galaksija}, we show the evolution of the compactness parameter 
$\Sigma_{1.5}$ as a function of the lookback time. Circles represent the main galaxy in the group (SubfindID 1287615),
while triangles and squares represent satellite galaxies, outliers from the mass-size relation (SubfindID 1287618 and 1287625). Vertical lines indicate the beginning and the end of the cluster phase.
The compactness of satellite galaxies steeply increases as soon as galaxies become a part of the group. Thus, in the group phase, galaxies already experience strong tidal stripping, leading to a decrease in the DM and gas components at the outskirts (Figure \ref{ID6627_masa}).
At the same time, new stars form at the core of the galaxy, stellar mass increases, and galaxies become more compact.
By the time the host group merges with a cluster, both galaxies have already reached a high level of compactness. This supports the idea that cEs are formed by tidal interactions with the massive host galaxy \citep{Deeley}.
Further cluster passage, as well as cluster merger, has only a slighter influence on the satellite galaxies' compactness parameter. On the other hand, the main galaxy becomes less compact during group and cluster phases.

\subsubsection{Possible morphological transformations of galaxies}
\label{morphology}

 A positive aspect of using the observational criteria defined to identify compact galaxies or the mass-size relation is that they rely on the global properties of galaxies (such as stellar mass and its half-mass radius). These global properties are reliably determined for systems that are sufficiently resolved to be detected accurately with structure finding algorithms, which typically require a few tens to a few hundred particles \citep{Onions+2012MNRAS.423.1200O}. Clearly, this is not an acceptable resolution for examining the internal structure of galaxies in more detail. This task requires at least a thousand stellar particles, whether we examine the structure of the galaxy manually \citep[e.g., by fitting the][profile]{Sersic1963} or use software for kinematical decomposition \citep[e.g., \textsc{Mordor} code developed by][]{Zana+2022MNRAS.515.1524Z} to determine the morphology of the galaxy. Almost all galaxies we examine in this work are below this particle threshold.

Fortunately, the IllustrisTNG simulation suite contains a supplementary catalogue for all simulation boxes, and for all galaxies that are sufficiently resolved, with information on stellar circularities and fractional stellar mass that is attributed to either spherical components, $F_\mathrm{Sph}$, or the thin disk, $F_\mathrm{Disk}$, \citep{Genel+2015ApJ...804L..40G}. We used this catalogue to check if the main galaxies of the three groups we examine in detail (ID6627, ID12982, ID15878) experience any morphological transformations during the cluster phase, comparing the fractional stellar mass of the two components before and after the cluster passage. The main galaxy in the ID6627 group has had $F_\mathrm{Disk} \simeq 0.34$ and $F_\mathrm{Sph} \simeq 0.33$ before the cluster infall, making it a likely late-type galaxy that transformed into a typical early-type galaxy post-cluster passage, with $F_\mathrm{Disk} \simeq 0.05$ and $F_\mathrm{Sph} \simeq 0.86$. Similarly, the main galaxy in the ID12982 group had $F_\mathrm{Disk} \simeq 0.18$ and $F_\mathrm{Sph} \simeq 0.42$ before the infall, which might not be a typical late-type galaxy. Still, with a generally low mass fraction of the spherical components and a moderately pronounced thin disk, one can consider it a lenticular galaxy. This galaxy also transformed into a typical early-type galaxy, with $F_\mathrm{Disk} \simeq 0.07$ and $F_\mathrm{Sph} \simeq 0.93$, after its group left the cluster environment. However, the main galaxy in the ID15878 group is not a typical early-type galaxy at present. With $F_\mathrm{Disk} \simeq 0.21$ and $F_\mathrm{Sph} \simeq 0.66$ in the present-day snapshot, the galaxy has a significant fraction of spherical components, but also a pronounced thin disk, which can make it either a lenticular galaxy or some intermediate type, but certainly not early or late-type exclusively. The catalogue does not contain information on its pre-cluster decomposition, since the galaxy did not have a sufficient number of stellar particles at the time; its stellar mass grew significantly during the cluster phase, when it experienced a peak in star formation. Both early-type galaxies at present (that have experienced morphological transformation) do not contain any gas. In contrast, the last galaxy we mentioned, which is not an early-type, still contains some gaseous component, with minimal star formation activity (star formation rate is about 0.11 $\mathrm{M}_\odot/\mathrm{yr}$).

Interestingly, the morphological transformations and the gaseous content (or lack thereof) do not appear to correlate with the number of accreted satellite galaxies. Instead, these scenarios appear to be directly influenced by the cluster passage itself and the pericentric distance of the orbit through the cluster. The fact that the main galaxy in the ID15878 group did not become a typical early-type galaxy, nor did it lose all of its gaseous content, should be easily and intuitively understood -- this group has only skimmed the outskirts of the cluster, as opposed to the two other groups. However, it should be clear that cluster passage can lead to morphological transformations and other significant changes in galaxy properties (e.g., quenching of star formation or a complete removal of gas through processes such as RPS). Unfortunately, examination of these scenarios in greater detail is beyond the scope of this work and typically requires higher resolution, which we intend to explore in the follow-up work.

\section{Conclusion}
\label{conclusion}

In this work, we investigated galaxy groups processed by a cluster, using IllustrisTNG300 cosmological simulation. 
Our focus was on groups with wide mass range from $8 \times 10^{11} \Msun$ to 
$7 \times 10^{13} \Msun$.
We distinguished between field groups that evolved in isolation and cluster groups that were part of a cluster during any period of their evolution. Cluster groups represent an example of rare objects ejected
from a cluster environment. Thus, studying it can provide insights into how the environment influences both groups as gravitationally bound systems, but also individual galaxies.
We further divided cluster groups into those that passed through a galaxy cluster and captured more
galaxies, referred to as thief groups, and groups that did not capture any new members, referred to as non-thief groups.

We found that thief groups are generally less compact and contain more members compared to field groups. Non-thief groups seem to have the same properties as the field groups, which makes them indistinguishable.
Employing different statistical tools to test the equality of field and cluster groups confirmed that field and thief groups do not belong to the same distribution (Table~\ref{stat_test}).
However, the cluster passage itself does not influence the group compactness, since non-thief groups follow the same distribution as field groups.

We further explored the evolutionary paths of the thief groups that captured a large number
($\sim7$) of new members and explored the influence of the passage through the cluster on individual galaxies (ID6627, ID12982, and ID15878).

We perform preliminary analyses of morphological transformations in those thief groups during their passage through a cluster. Due to simulation resolution limits, we focus on the main galaxy in each group. The results indicate that the groups passing within the cluster's virial radius show transformations toward early-type morphologies, while the group that only skimmed the outskirts retains gas and a moderate disk component. These findings indicate that morphological transformation and gas removal are more closely correlated with the pericentric distance of the group's orbit through the cluster than with the number of accreted galaxies.
In this study, we choose a large volume TNG300 simulation which provides a statistically significant sample of galaxy clusters. However, this comes at the cost of lower resolution, restricting our ability to carry out detailed morphological analyses of individual galaxies within the groups. A more detailed investigation of galaxy morphology will be pursued in a follow-up study using a higher-resolution simulation, such as TNG100, though this would necessarily involve a smaller sample of thief groups.

Our sample contains one thief group with two satellite galaxies that are outliers from the galaxy mass-size relation, being more compact than expected for the given mass.
We explored the evolution of this group and showed that satellite galaxies become compact in the group phase
before the cluster infall, as a result of tidal interactions with the massive host galaxy.
This finding is in agreement with a recent study by \citet{Deeley}.
Even though the further evolution of this group takes an unusual path where it makes a passage within a cluster virial radius, experiences a cluster-cluster merger, and then escapes the cluster while accreting new galaxies, the compactness parameter of satellite galaxies does not change significantly in the cluster phase. 

This study offers a first step toward identifying and characterizing galaxy groups that have been processed by clusters and later escaped, potentially capturing new members in the process. Our results suggest that such thief groups may be observationally identifiable as field groups with unusually high membership and low compactness for their mass. Future observational work could focus on confirming these signatures through deep spectroscopic and morphological surveys, particularly targeting diffuse groups near massive clusters or along filamentary structures. On the theoretical side, higher-resolution simulations will be essential to resolve the internal galaxy structure and better capture morphological transformations, gas stripping, and dynamical interactions within these groups. Together, these approaches will refine our understanding of the long-term impact of cluster environments on group and galaxy evolution.


\paragraph{Funding Statement}

This research was supported by the Ministry of Science, Technological Development and Innovation of the Republic of Serbia (MSTDIRS) through contract no. 451-03-136/2025-03/200002 made with Astronomical Observatory of Belgrade.

\paragraph{Competing Interests}

None
\paragraph{Data Availability Statement}

The data generated in this study are available from the corresponding author, M.S., upon reasonable request.

\paragraph{Ethical Standards}
The research meets all ethical guidelines, including adherence to the legal requirements of the study country.

\paragraph{Author Contributions}

Conceptualization: M.M.; M.S.; A.M.
Formal analysis: M.S.; A.M.
Methodology: M.S.; M.M.; A.M.
Visualization: M.S.; A.M.
Writing original draft: M.S.
Writing review \& editing: A.M.; M.M. All authors discussed the results and contributed to the final version of the manuscript. All authors approved the final submitted draft.

\printendnotes

\printbibliography

\end{document}